\documentclass[11pt,a4paper]{scrartcl}

\usepackage{ILD}

\usepackage[symbol]{footmisc}
\usepackage{feynmf}
\usepackage{multirow}
\usepackage{textpos}
\usepackage{lineno}
\usepackage{afterpage}

\usepackage{tabularx}
    \newcolumntype{L}{>{\raggedright\arraybackslash}X}

\def\kaonness{\ensuremath{\Delta_{\dEdx-K}}\xspace}

\def\epsilonb{\ensuremath{\epsilon_{b}}\xspace}
\def\epsilonc{\ensuremath{\epsilon_{c}}\xspace}
\def\epsilonuds{\ensuremath{\epsilon_{uds}}\xspace}
\def\epsilonb2{\ensuremath{\epsilon^{2}_{b}}\xspace}
\def\epsilonc2{\ensuremath{\epsilon^{2}_{c}}\xspace}
\def\epsilonuds2{\ensuremath{\epsilon^{2}_{uds}}\xspace}

\def\costheta{\ensuremath{\cos \theta}\xspace}

\def\fb{fb\ensuremath{^{-1}}\xspace}

\def\bbbar{\ensuremath{b}\ensuremath{\overline{b}}\xspace}
\def\qqbar{\ensuremath{q}\ensuremath{\overline{q}}\xspace}
\def\ccbar{\ensuremath{c}\ensuremath{\overline{c}}\xspace}

\def\eeqqbar{\ensuremath{e^{-}e^{+}\rightarrow q\bar{q}}\xspace}

\def\eeZgammaqqbar{\ensuremath{e^{-}e^{+}\rightarrow Z \gamma \rightarrow q\bar{q} \gamma}\xspace}

\def\eebb{\ensuremath{e^{-}e^{+}\rightarrow b\bar{b}}\xspace}
\def\eecc{\ensuremath{e^{-}e^{+}\rightarrow c\bar{c}}\xspace}
\def\eeqq{\ensuremath{e^{-}e^{+}\rightarrow q\bar{q}}\xspace}

\def\eLpR{\ensuremath{e_L^{-}e_R^{+}}\xspace}
\def\eRpL{\ensuremath{e_R^{-}e_L^{+}}\xspace}

\def\dEdx{\ensuremath{dE/\/dx}\xspace}
\def\dNdx{\ensuremath{dN/\/dx}\xspace}

\def\Afb{\ensuremath{A_{FB}}\xspace}
\def\AFB{\ensuremath{A_{FB}}\xspace}

\def\Rq{\ensuremath{R_{q}}\xspace}

\def\bquark{\ensuremath{b}-quark\xspace}

\def\bjets{\ensuremath{b}-jets\xspace}

\def\btag{\ensuremath{b_{tag}}\xspace}
\def\cquark{\ensuremath{c}-quark\xspace}

\def\cjets{\ensuremath{c}-jets\xspace}

\def\ctag{\ensuremath{c_{tag}}\xspace}

\def\udsjets{\ensuremath{uds}-jets\xspace}

\def\Bc{\ensuremath{Vtx}-method\xspace}
\def\Kc{\ensuremath{K}-method\xspace}

\def\Pb{\ensuremath{P_{chg.}}\xspace}

\title{Experimental prospects for precision observables in \eeqq with $q=b,c$ processes at the ILC operating at 250 and 500 GeV of center of mass.}
\titlecomment{This work was carried out in the framework of the ILD concept group}
\titlecomment{Talk presented at the International Workshop on Future Linear Colliders (LCWS 2023), 15-19 May 2023. C23-05-15.3.}

\ildproc{PHYS}{2023}{004, IFIC/23-37} 

\date{\today}

\addauthor{A. Irles}{\institute{1} \footnote{\href{mailto:adrian.irles@ific.uv.es}{adrian.irles@ific.uv.es}}}
\addauthor{J.P. M\'arquez}{\institute{1} 
\footnote{Speaker \href{mailto:jesus.marquez@ific.uv.es}{jesus.marquez@ific.uv.es}}}

\addinstitute{1}{IFIC, Universitat de Val\`encia and CSIC, C./ Catedr\'atico Jos\'e Beltr\'an 2, E-46980 Paterna, Spain}

\abstract{
Future Higgs Factories will allow the precise study of \eeqq with $q=s,c,b,t$ interactions at different energies, from the Z-pole up to high energies never reached before.
In this contribution, we will discuss the experimental prospects for the measurement of differential observables in \eebb and \eecc processes at high energies, 250 and 500 GeV, using full simulation samples and the full reconstruction chain from the ILD concept group.
These processes call for superb primary and secondary vertex measurements, a high tracking efficiency to correctly measure the vertex charge and excellent hadron identification capabilities using \dEdx. This latter aspect will be discussed in detail together with its implementation within the standard flavour tagging tools developed for ILD (LCFI+). In addition, prospects associated with potential improvements using cluster counting techniques instead of traditional \dEdx will be discussed. 
}

\graphicspath{ {./logos/}{./figures/} }

\begin{document}

\titlepage

\tableofcontents

\section{Introduction}
\label{sec:intro}

The Standard Model (SM) was successfully confirmed when the Higgs boson was discovered by the Large Hadron Collider (LHC)\cite{Aad_2012,Chatrchyan_2012} experiments. Since then, the measurement of its properties with the highest accuracy has become a primary goal.
For that reason, in parallel with the LHC data exploitation, the high-energy accelerator-based particle physics community works towards constructing the next large accelerator after the LHC. Such a machine will be an $e^{+}e^{-}$ collider at relatively high energy to achieve the highest precision on the Higgs physics measurements. 
Different projects have been proposed and are under discussion.
These are the so-called \textit{Higgs factories}.

The International Linear Collider (ILC)\cite{Behnke:2013xla,Baer:2013cma,Adolphsen:2013jya,Adolphsen:2013kya,Behnke:2013lya} is one of the \textit{Higgs factories} proposed for the future. 
The ILC operation foresees $e^{+}e^{-}$ collisions at different centres of mass energies ranging from the Z-pole (Giga-Z) to up to 1 TeV after energy upgrades. 
The nominal program defines an initial stage at 250GeV (ILC250), with a luminosity upgrade, and a following one at 500GeV (ILC500), after an energy upgrade. 
The ILC also features polarised beams: $80\%$ for electrons and $30\%$ for positrons.
The ILC250 physics program foresees a total integrated luminosity of 2000 \fb distributed between four different beam polarisation schemes: $45\%$ in $P_{\mathrm{e^{-}e^{+}}}=(-0.8,+0.3)$, $45\%$ in $P_{\mathrm{e^{-}e^{+}}}=(+0.8,+0.3)$, $5\%$ in $P_{\mathrm{e^{-}e^{+}}}=(-0.8,-0.3)$ and $5\%$ in $P_{\mathrm{e^{-}e^{+}}}=(+0.8,-0.3)$. 
The ILC500 program foresees a total integrated luminosity of 4000 \fb distributed among the different polarisation configurations in the same proportions as the ILC250 case.

The ILD~\cite{Behnke:2013lya,ILD:2020qve} is one of the two proposed detector concepts for the ILC. 
It is a highly-hermetic multi-purpose detector designed for the maximal exploitation of particle flow techniques in event reconstruction. 

This document updates the studies presented in \cite{Irles:2023ojs}.
In that work, the experimental prospects for \eeqq measurements at ILC250 are studied. 
The study of such topologies is critical to obtain a complete picture of the electroweak sector interactions between SM bosons and fermions, particularly quarks.
For instance, models of new physics featuring extra-dimensions \cite{Yoon:2018xud,Funatsu:2017nfm,Funatsu:2020haj} have been proposed to explain the striking mass hierarchy in the fermion sector. These models predict deviations in the electroweak sector with  modification of the SM couplings and new heavy resonances $Z^{\prime}$ that interact with the fermions.
Furthermore, the LEP/SLC anomaly in \eebb is still unexplained \cite{Djouadi:2006rk}. The effects of new physics may differ for different fermion chiralities and additional terms associated with various mediators (SM $Z$ and $\gamma$ or beyond SM $Z^{\prime}$ or mixing of these). This motivates the study of quark pair production in high energy $e^{-}e^{+}$ collisions at past lepton colliders \cite{ALEPH:2005ab} and at future ones \cite{Irles:2023ojs}.  

In \cite{Irles:2023ojs}, two experimental observables have been studied in detail for the ILC250 using the ILD detector in full simulation.
The observables are the Hadronic Fraction (\Rq) and Forward-Backward Asymmetry (\AFB). The observable \Rq is defined as:
\begin{equation}
R_{q}=\frac{\sigma_{e{-}e^{+}\rightarrow q\bar{q}}}{\sigma_{had.}}
\label{formula:Rq}
\end{equation} 
where $\sigma_{had.}$ is the integration of $\sigma_{q\bar{q}}$ for all quark flavours except the top quark. In the case of \AFB the definition reads:
\begin{equation}
A^{q\bar{q}}_{FB}=\frac{\sigma^{F}_{e{-}e^{+}\rightarrow q\bar{q}}-\sigma^{B}_{e{-}e^{+}\rightarrow q\bar{q}}}{\sigma^{F}_{e{-}e^{+}\rightarrow q\bar{q}}+\sigma^{B}_{e{-}e^{+}\rightarrow q\bar{q}}}
\label{formula:AFB}
\end{equation}
where $\sigma^{F/B}_{e{-}e^{+}\rightarrow q\bar{q}}$ is the cross-section in the forward/backward hemisphere as defined by the polar angle $\theta_q$.

In this contribution, we report on incorporating an improved flavour tagger that uses the hadron identification (protons, kaons, pions) as input. In addition, the study has been extended to include the prospects of the same study at ILC500. Finally, we include a prospective study of the precisions that could be reached if a pixel TPC is used for the ILD.

\section{The ILD Concept and MC Samples}
\label{sec:ILD}

The ILD subdetector layout consists of a high-precision vertex detector (VTX), silicon tracking systems, a time projection chamber (TPC), a highly granular calorimeter system (ECAL and HCAL) and a muon catcher. All the aforementioned subdetectors are placed inside a solenoid providing a magnetic field of $3.5$\,T, surrounded by an iron yoke instrumented for muon detection. The ILD TPC \cite{thelctpccollaboration2016time,LCTPC:2022pvp} is a large volume time projection chamber that allows continuous 3D tracking and particle identification. 
Its baseline design consists of a barrel-shaped structure with an inner radius of 329 mm and an outer radius larger than 1808 mm. 
It provides a single-point resolution of 100 \textmu m over about 200 readout points and a \dEdx resolution of $\sim4.5\%$. It is based on $1\times 6$ mm$^{2}$ pads readout with GEM or MicroMegas technologies.
This document will also explore the potential of an alternative approach: a pixel TPC. Several technological solutions are under study, and all envision a four times greater readout density with a pixel size of 300 \textmu m. Simulations extrapolating beam test results show that an improved relative resolution of $\sim3$-$4\%$ will be feasible using cluster counting techniques (\dNdx) instead of the traditional approach \cite{LCTPC:2022pvp}. 

This study has been conducted running full simulation via \texttt{ILCSOFT} \textit{v02-02-03}\footnote{Link: \url{https://github.com/iLCSoft}}, which merge different software packages and algorithms that operate in a modular way from MC events to final reconstruction. All simulations use the ILD-L\cite{ILD:2020qve} model, whose geometry, material and readout systems are implemented in the \texttt{DD4HEP} framework\cite{Frank:2014zya}, interfaced with \texttt{Geant4} toolkit. Both signal and background events are generated with the \texttt{WHIZARD}\cite{Kilian:2007gr} event generator at LO. The beam energy spectrum, beam-beam interaction and QED ISR is generated via \texttt{Guinea-Pig}\cite{Schulte:1999tx}. Non-perturbative effects, such as the FSR and hadronization, are provided by \texttt{Pythia} event generator\cite{Sj_strand_2006}.

To reproduce different stages of the ILC operation, two different samples have been provided by the ILD concept group; one for 250 GeV and one for 500 GeV. Each set of samples features full polarised beams in different configurations. The main samples for this study are the $\mathrm{e^-_{L}}\mathrm{e^+_{R}}$ and $\mathrm{e^-_{R}}\mathrm{e^+_{L}}$. Background processes from electroweak bosons, Higgs and top-quark production (at 500 GeV) are studied using additional MC samples. Backgrounds from the production of lepton pairs have been ignored since they are expected to be easily identified.
All distributions and results have been reweighted to match the baseline polarization and full luminosity scenarios for ILC250 and ILC500, so-called the H20 scenario \cite{Bambade:2019fyw}.

The generation of the \eeqqbar signal events and the \eeZgammaqqbar with the Z-boson generated on-shell are done simultaneously. Following the same recipe as in \cite{Irles:2023ojs}, we define the signal as the events with acolinearity of the \qqbar system (at parton level) smaller than 0.3 and a large invariant mass of the \qqbar pair larger than 140 GeV - for the ILC250 case - or larger than 200 GeV for the ILC500 case  (i.e. large enough to be away of the Z-mass peak).

The signal and background cross sections are listed in Tables 1 and 2 from \cite{Irles:2023ojs} for ILC250 and in Tables \ref{tab:crosssection500} and \ref{tab:crosssection_bkg500} for ILC500. Only backgrounds leading to fully hadronic final states are considered. Backgrounds involving leptons in the final states are ignored since those are expected to be easily identified.

\begin{table}[!ht]
  \centering
  \begin{tabular}{c|ccc|ccc}
    \hline
     & \multicolumn{3}{|c|}{ $\sigma_{\eeqqbar}$[fb]} & \multicolumn{3}{|c}{ Radiative Return BKG [fb]} \\
    \hline
    Polarisation & \bbbar & \ccbar & \qqbar ($q=uds$) & \bbbar & \ccbar & \qqbar ($q=uds$) \\
    \hline
    \eLpR & 611.4 & 1545.9 & 2770.2 & 5506.4 & 5118.0 & 16244.2\\
    \eRpL & 416.9 & 893.8 & 1728.3 & 3002.5 & 2766.9 & 8773.5 \\
    \hline
  \end{tabular}
  \caption{Production cross section of signal and background originated by di-boson production. \label{tab:crosssection500}}
\end{table}

\begin{table}[!ht]
 \centering
  \begin{tabular}{c|c|c|c|c}
    \hline
    & \multicolumn{4}{|c}{  $\sigma_{e^{-}e^{+}\rightarrow\,X}$ [fb]} \\
    \hline
	Polarisation &  $X= WW \rightarrow q_{1} \bar{q_{2}} q_{3} \bar{q_{4}}$ & $X= ZZ \rightarrow q_{1} \bar{q_{1}}q_{2} \bar{q_{2}}$ & $X= HZ \rightarrow q\bar{q}H$ & $X=t\bar{t} \rightarrow b\bar{b}q_{1}\bar{q}_{2}q_{3}\bar{q}_{4}$\\
    \hline
    \eLpR & 7680.0 & 680.2 & 114.7 & 660 \\
    \eRpL & 33.5 & 271.9 & 73.4 & 254.8 \\
      \hline
  \end{tabular}
  \caption{\label{tab:crosssection_bkg500} Cross sections at 500 GeV for processes producing at least one pair of $q$-quarks, with fully polarised beams. }
\end{table}

\section{\Rq and \AFB reconstruction at ILD}
\label{sec:recosel}
The reconstruction of the events begins with the track reconstruction performed by the \texttt{MarlinTrk} framework, part of \texttt{ILCSoft}. 
Later, \texttt{Pandora}\cite{Marshall:2015rfa} runs the particle flow algorithm (PFA) that matches the tracking information with the high-granular calorimetry information following pattern recognition techniques. The resulting reconstructed objects are denominated particle flow objects (PFO) and are treated as single particles.

Once the PFOs are reconstructed, the vertex reconstruction, jet reconstruction and flavour tagging are performed with \texttt{LCFI+} software tool.

Once all high-level reconstruction objects are obtained, the event preselection is performed. It is based on a series of kinematical cuts applied to enrich the data sample with signal events while removing the backgrounds. The preselection for 250 GeV is the same as described in \cite{Irles:2023ojs}, where the variables are described, with minor adaptations made for the study of the 500 GeV samples (noted in parenthesis): 
\begin{enumerate}
    \item photon veto cuts, rejecting events if:
    \begin{enumerate}
        \item at least one of the jets contains only one PFO;
        \item at least one of the jets contains a reconstructed $\gamma_{cluster}$ with E > 115 GeV (220 GeV) or located in the forward region $|\costheta|>0.97$;
    \end{enumerate}
    \item events with $\sin \Psi_{acol}$ > 0.3 are rejected;
    \item events with $m_{jj}$ < 140 GeV (200 GeV) are rejected;
    \item events with $y_{23}$ > 0.02 (0.007) are rejected.
\end{enumerate}
The most noticeable change for the ILC500 case is performed in the last cut, which is tightened to reduce the much larger $WW$ background contamination in the case of left-handed polarisation. This last cut decreases the final preselection efficiency to $\sim50\%$ instead of $\sim75\%$ for the ILC250 case.

The next step is applying the Double Tag (DT) method stated in \cite{Irles:2023ojs} by which a cut is performed in the \btag (or \ctag) likelihood in both jets of the event. The working points are defined as in \cite{Irles:2023ojs} such that the mistagging of the other quarks when performing the tagging of \bquark (or \cquark) is smaller than the $\sim1.5\%$ $(\sim 3\%)$, assuming the same size for the $b/c/uds$ samples. The DT yields a high-purity selection of the events for each flavour and is the last necessary step to reconstruct $R_q$ measurements.

To study \AFB, the charge of the jets needs to be reconstructed to fully determine the direction of the $q\bar{q}$ system. As described in \cite{Irles:2023ojs}, two independent methods are used for the charge measurement: \Bc or \Kc. The \Bc counts the charge of all tracks in secondary vertices in the jet, while the \Kc only uses the information of tracks identified as Kaons from the TPC dE/dx PID. The Double Charge (DC) method performs the final charge measurement, which combines the possible measurements of both methods and accepts only jets with opposite charges. 
The differential, in \costheta, cross section is reconstructed, and the \AFB extracted from a fit restricted to ($\|\costheta\|>0.9$) to avoid the region where the reconstruction efficiency drops due to insufficient acceptance of the tracking detectors. 

All the methods (Preselection, DT, DC, Fit) have been developed for the analysis at 250 GeV and are explained in detail in \cite{Irles:2023ojs} together with a comprehensive study of the most dominant systematic uncertainties.
The same methods are used for the study at 500 GeV shown in this document.
In Appendix \ref{sec:appendixPlots}, we collect the most relevant performance plots for the ILC500 case to be compared with the ones in \cite{Irles:2023ojs}.


\section{Optimisation of the $\Afb$ reconstruction at ILC}
\label{sec:optimization}
\subsection{Optimisation of the default flavour tagging of ILD: Adding hadron PID as input.} \label{PIDsection}

The flavour tagging is performed by the \texttt{LCFI+} package \cite{Suehara:2015ura}.
The vertex reconstruction method distinguishes between vertices and pseudo-vertices, which are \textit{single-track vertices} candidates to be merged in a fully reconstructed vertex in a second iteration of the algorithm. At the end of the process, if these pseudo-vertices are not merged in a full vertex, they are kept in the list of objects and treated as \textit{single-track vertices}.

The total number of vertices and pseudo-vertices is always two or fewer. Four different categories are defined (A, B, C, D): A are jets without vertices and up to two pseudo-vertices, B are jets with one vertex and no pseudo-vertices, C are jets with one vertex, and one pseudo-vertex and D are jets with two vertices. The flavour content in each category is different as expected due to the different hadronisation products of heavy quarks: Above 95\% of light-quark jets are in category A, more than 90\% of \cquark jets are spread between categories A and B while more than 80\% of the \bquark jets are split between categories B, C and D.

To prepare weights for flavour tagging via \texttt{LCFI+}, two main algorithms have to be run: \textit{MakeNTuple} and \textit{TrainMVA}. MakeNTuple prepares the \texttt{ROOT} files with NTuples with all the different variables that could be used for the flavour tagging. TrainMVA uses 3 NTuples as input (one for \bjets, one for \cjets and one for \udsjets) and then runs a classificator based on Boosted Decision Trees (BDT implemented in \texttt{ROOT}'s TMVA). The input is the signal data separated into the different categories described above. A \texttt{Marlin} processor\footnote{Link: \url{https://github.com/marherje/LCIO_Extraction.git}} was designed \textit{ad hoc} to do this selection. 

New variables were built using the TPC's PID capabilities via \dEdx to improve the flavour tagging. The new variables were a count of kaons, protons and pion candidates. The charged kaon, proton and pion IDs are defined by constructing \textit{likenesses} variables using a statistical distance measurement, comparing theoretical and experimental values. for instance, the definition of \textit{kaonness}, \kaonness, is defined as:
\begin{equation}
\Delta_{\dEdx-K}=\left(\frac{dE/dx_{exp}-dE/dx_{K,BB}}{\Delta dE/dx_{exp}}\right),
\label{formula:kaonness}
\end{equation}
where $dE/dx_{K,BB}$ is the theoretical value given by the Bethe-Bloch formula, $dE/dx_{exp}$ is the experimental value from the TPC's measurements and $\Delta dE/dx_{K,exp}$ is the error associated to that measurement as implemented in the official ILD reconstruction software. This definition is extended to other hadron types (\textit{pionness}, \textit{protonness}). This distance can be modelled as normally distributed. It can be constructed concerning each particle's theoretical values, hence having the Gaussian centred in 0 for each case (\textit{kaonness}, \textit{pionness}, \textit{protonness}). To have a proper distribution, all tracks with momenta below 3 GeV and $\|\cos\theta\| > 0.95$ were discarded before selecting particle candidates. Low-momenta tracks are removed because the Bethe-Bloch curves at low energies for pions and kaons overlap. Removal of very forward/backward tracks is for a loss of geometrical acceptance of the TPC. We count as candidate particles of each type those whose absolute \textit{likeness} value is below 1.5, which minimises overlapping with other particles. All these new variables and selections were implemented in a beta version of \texttt{LCFI+} that is already available and waiting for an official \textit{pull} to the main \texttt{GitHub} repository of \texttt{LCFI+}. This process counts only tracks associated with a secondary vertex (or pseudo-vertex).
The final three new variables added to \textit{MakeNTuple} are named: dEdxNKaonSec, dEdxNProtonSec and dEdxNPionSec, as they represent the number of each particle in secondary vertices.

There is a direct correlation between quark flavour and kaon production via B-mesons (\bquark) and D-mesons (\bquark and \cquark) decays which motivate using kaon PID in flavour tagging. Identification of kaons is also useful for the charge reconstruction as stated in Sec. \ref{sec:recosel}, which is especially important in the case of \cquark. Using \dEdx is strongly motivated given the expected separation power of high momenta tracks (above 3 GeV)(see \cite{ILD:2020qve}, Figure 8.6). 

The standard training process runs once through the four vertexing categories using the same configuration of the BDTs. Still, in the case of this optimisation, independent training has been performed for each category. To ensure that the BDT configuration is optimal for each different scenario (energy, category and variable selections), a Particle Swarm Optimisation (PSO) was run. 
The PSO\cite{488968} is a parameter-free, stochastic, bio-inspired algorithm that searches for the optimal configuration of a problem by doing an iterative scanning of configurations for the problem.
The \textit{particles} are positions in the space of configurations for the given problem, which move to a new position after each iteration.
It requires the definition of a Function Of Merit (FOM) as scoring to determine the movement of the \textit{particles} after each iteration. 
In this case, a PSO software\cite{CMS:2018hnq} has been adapted and extended to a 3-class classifier with filters suitable to work with \texttt{LCFI+}\footnote{Link: \url{https://github.com/marherje/PSOforLCFIPlus.git}}. The space of configurations for flavour tagging are the hyper-parameters of the BDTs, and the \textit{particles} of the PSO are different selections of such hyper-parameters. The different hyper-parameters for the BDTs are the number of trees, the maximum number of leaves for each tree, the shrinkage, the bagging fraction and the number of bins (for the physical input variables' histograms). The function of merit of choice for this study was the average value of the integral of the Receiving Operating Characteristic (ROC) curve for the three different flavour categories ($b$, $c$, $uds$) from the test sample. To avoid overfitting to the test sample, two different statistical tests (Kolmogorov-Smirnov\cite{kolmogorov1933sulla,smirnov1948table,hodges1958significance} and Anderson-Darling\cite{cf37c5fc-d933-3771-9586-9d6b4b285d8b,doi:10.1080/01621459.1954.10501232,engmann2011comparing}) were run each time a particle finished its training, comparing the BDTs output between the test and training samples. In both statistical tests, a threshold value was set upon observation, always keeping the p-value over the standard $p>0.05$ criteria but also checking that the fluctuations in the FOM values were below 1.5 \%. The complete list of variables used in each category is in Table \ref{tab:bdtvariables} of the Appendix \ref{sec:annex}, where the bold text are the new variables and the rest are original \textit{LCFI+} variables as introduced in \cite{Suehara:2015ura}.

\subsection{Prospects of using a pixel TPC device: From \dEdx to \dNdx } \label{dNdxsection}
A better PID resolution is expected when using \dNdx instead of \dEdx. This will be possible with a pixel-based TPC. 
According to \cite{LCTPC:2022pvp}, an improvement of $\sim 30-40\%$ in the kaon/pion separation power is achievable for tracks with momentum between 3-50 GeV, with the separation power as defined in  \cite{ILD:2020qve} and in the following equation: 
\begin{equation}
    \eta_{A,B}(p)=\frac{\lvert\mu_A(p)-\mu_B(p)\rvert}{\sqrt{\frac{1}{2}(\sigma_A^2(p)+\sigma_B^2(p)}},
    \label{eq:separationpower}
\end{equation}
where $\mu_(p)$ and $\sigma$ are the mean and standard deviations of the gaussian fits to the experimental value of \dEdx, defined as a function of the momenta $p$ of each bin. 
However, the \dNdx reconstruction is not yet implemented in the ILD software, so it is not possible to introduce a \dNdx PID variable for the flavour tagging and charge measurement directly, as for \dEdx. 
Instead, we apply a correction to the \textit{likenesses} distributions by the expected value.
This is done by a dedicated Marlin processor that emulates a $25\%$ reduction in standard deviation in the 
\textit{likenesses} distributions.
This $25\%$ reduction in the  standard deviation in the 
\textit{likenesses} distributions is equivalent to a to a $\sim33\%$ improvement in the separation power,
In addition, in the flavour tagging case, an Ideal PID resolution has been considered by reducing the standard deviation of the simulated \textit{likeness} distributions by a $99\%$, compared with the full simulation.

\begin{figure}
    \centering
    \includegraphics[width=0.45\textwidth]{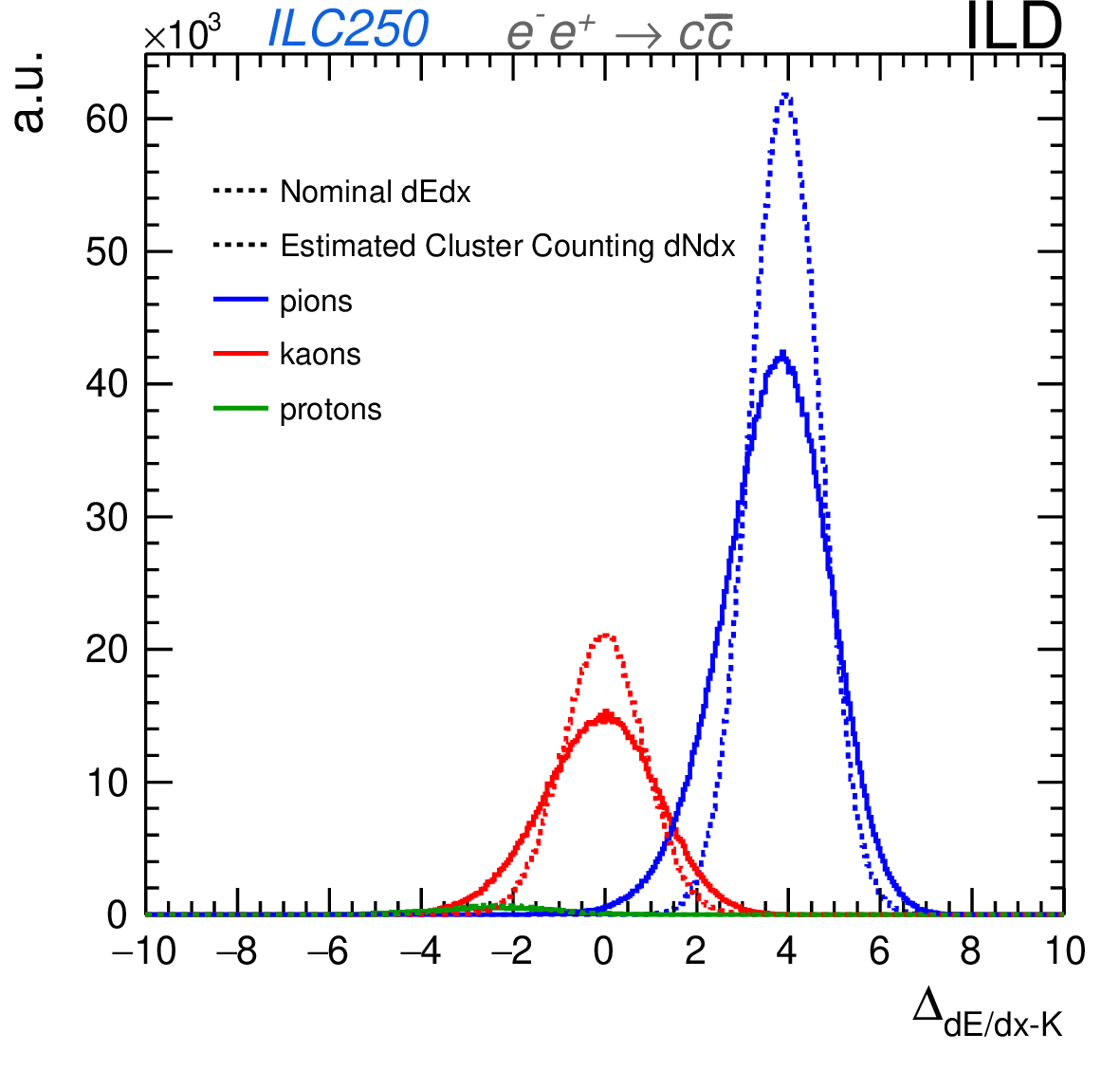}
    \caption{Kaonness distribution of secondary tracks in c-quark jets at ILC250 for protons, kaons and pions. The continuous line represents the real \dEdx distribution, while the dotted line represents the \dNdx estimation.}
    \label{fig:dEdxdist_cquark}
\end{figure}

\begin{figure}[!ht]
\begin{center}
    \begin{tabular}{cc}
      \includegraphics[width=0.45\textwidth]{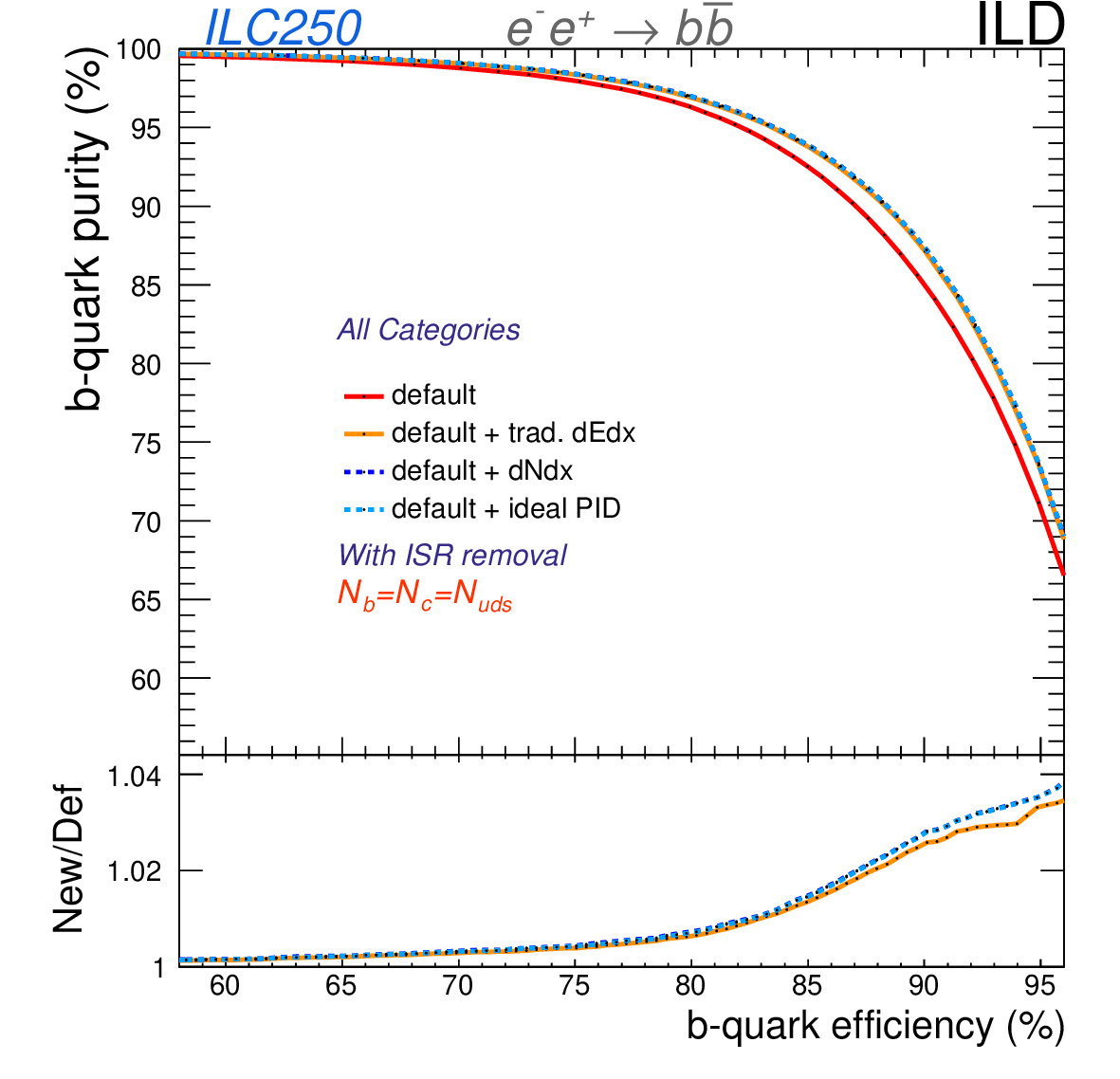} &
      \includegraphics[width=0.45\textwidth]{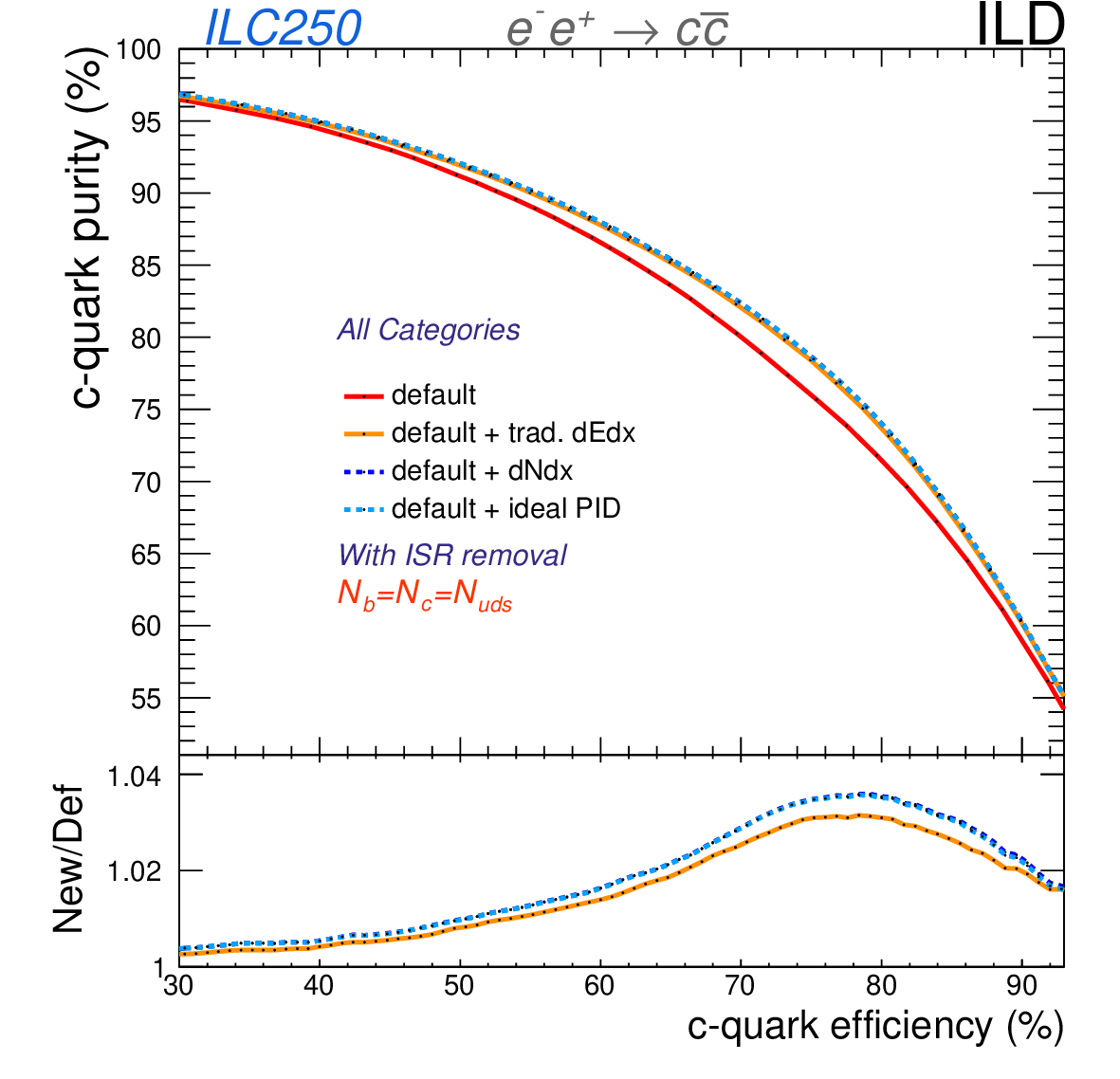} \\
      \includegraphics[width=0.45\textwidth]{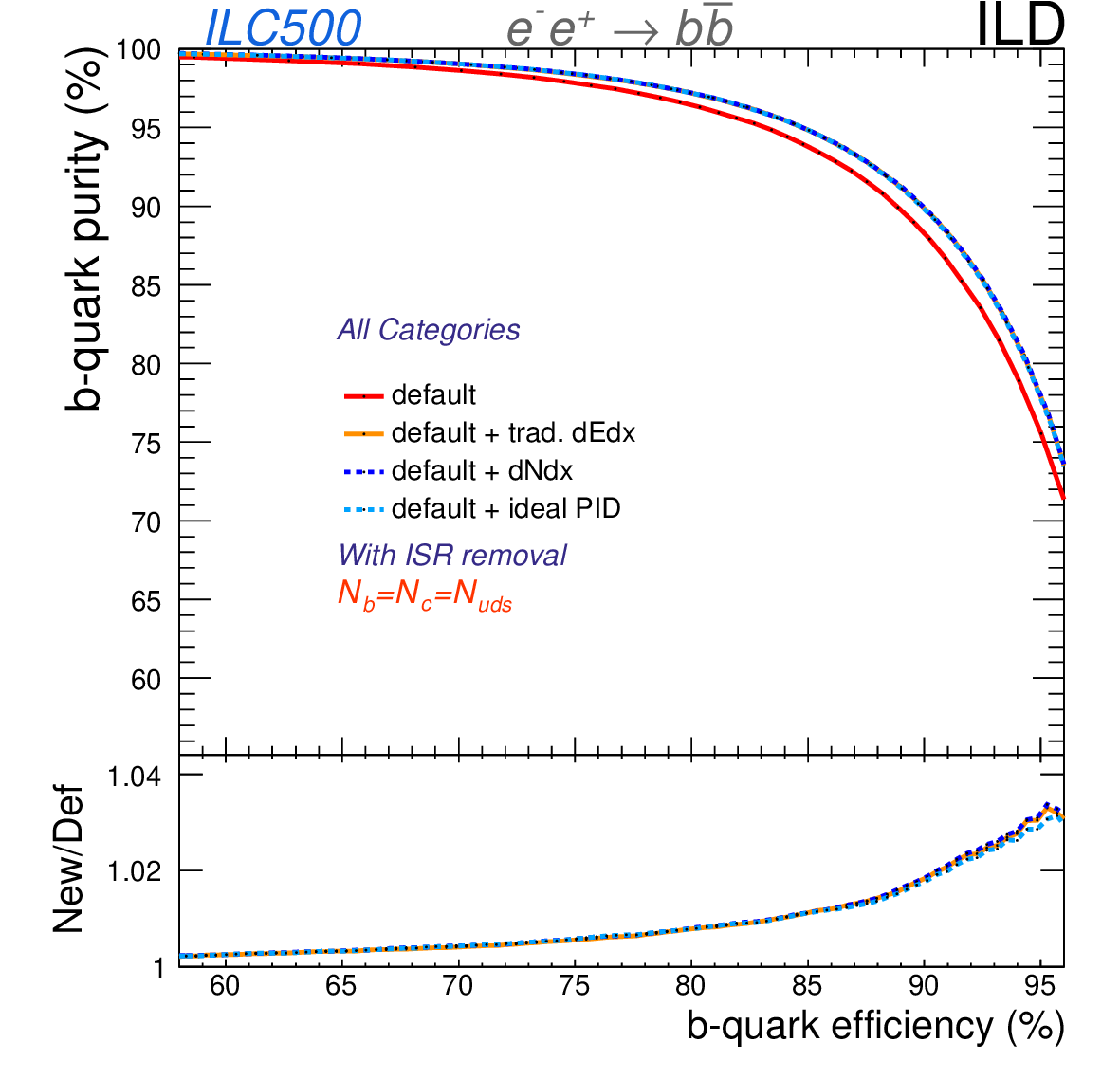} &
      \includegraphics[width=0.45\textwidth]{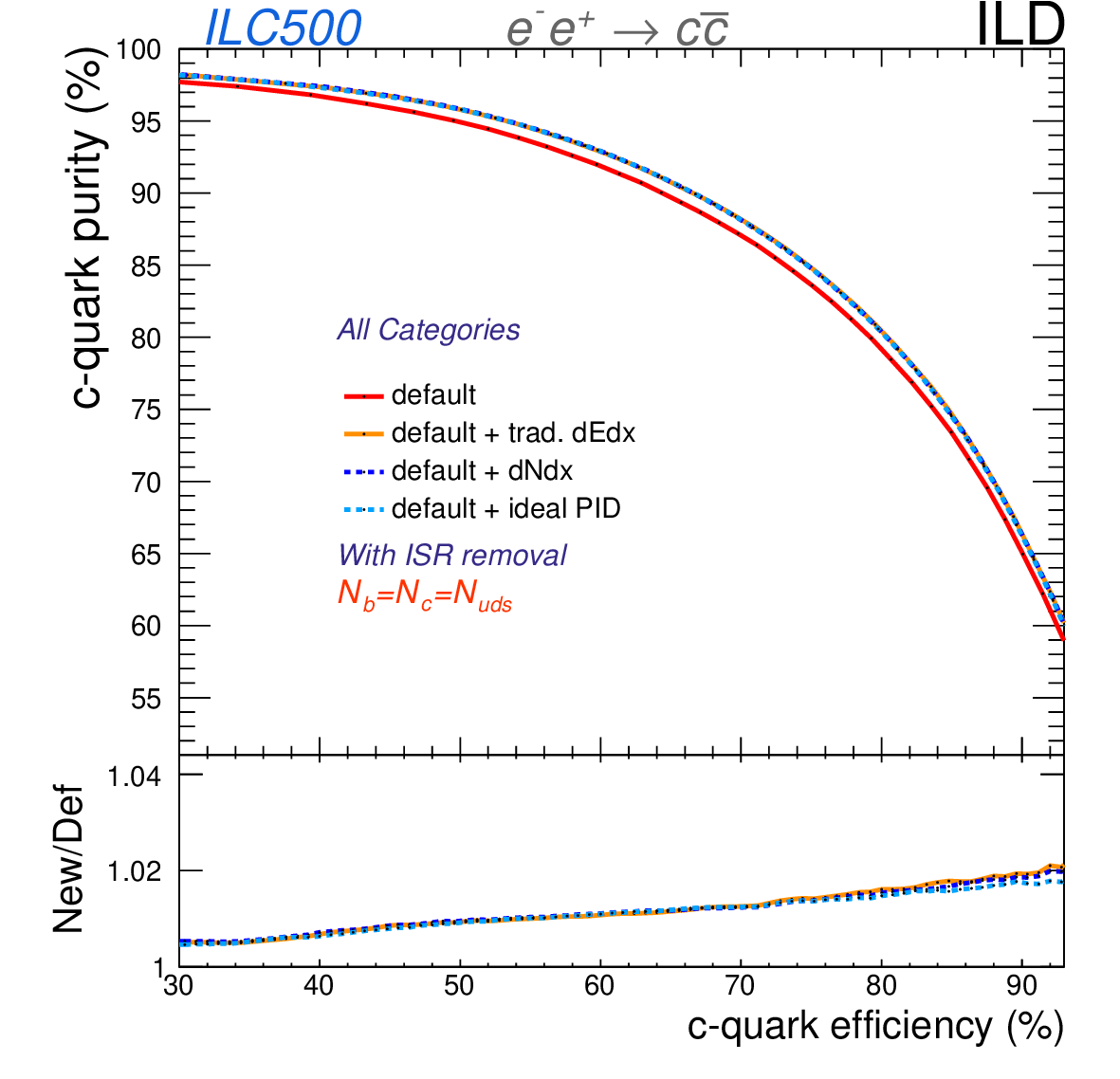} 
    \end{tabular}
\caption{Prospects on flavour tagging (purity vs efficiency). On top row: effects of b-tagging in the left plot and c-tagging in the right plot, both at 250 GeV. The second row shows the result for 500 GeV. The panel below represents the ratio between the new taggings and the default selection on all plots. \label{fig:flavourtagging}}
\end{center}
\end{figure}

\begin{figure}[!ht]
\begin{center}
    \begin{tabular}{cc}
      \includegraphics[width=0.45\textwidth]{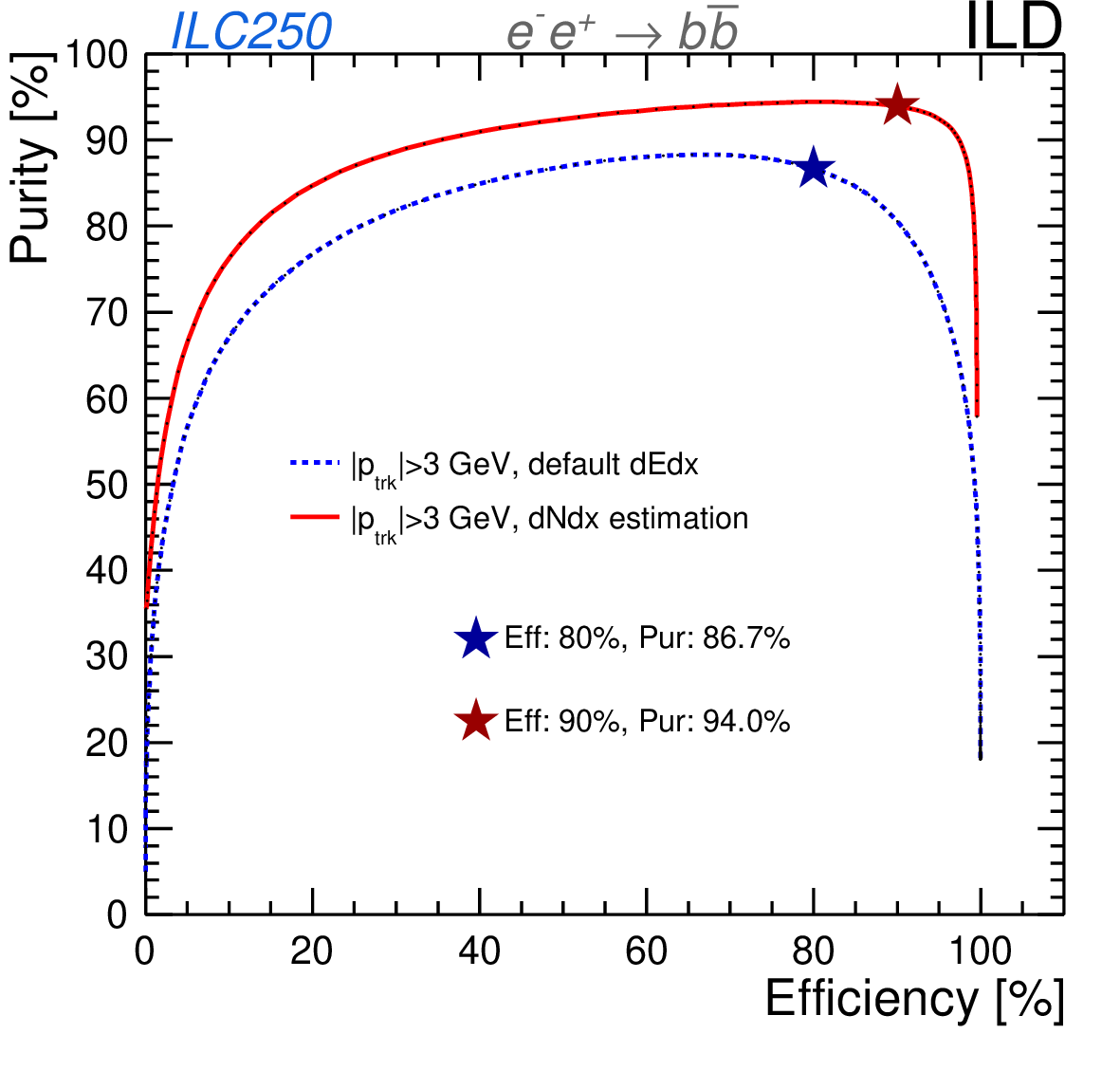} &
      \includegraphics[width=0.45\textwidth]{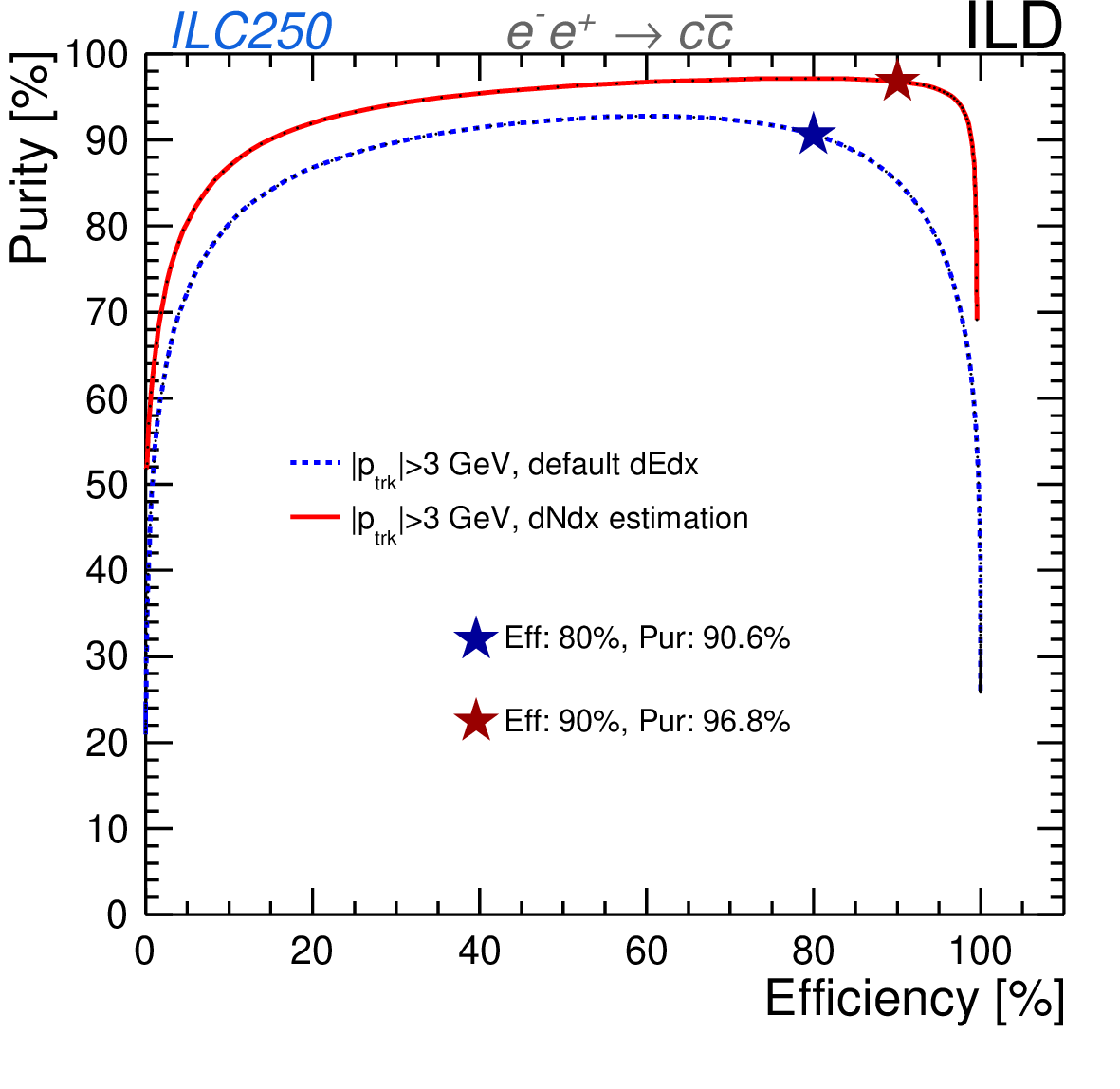} \\
      \includegraphics[width=0.45\textwidth]{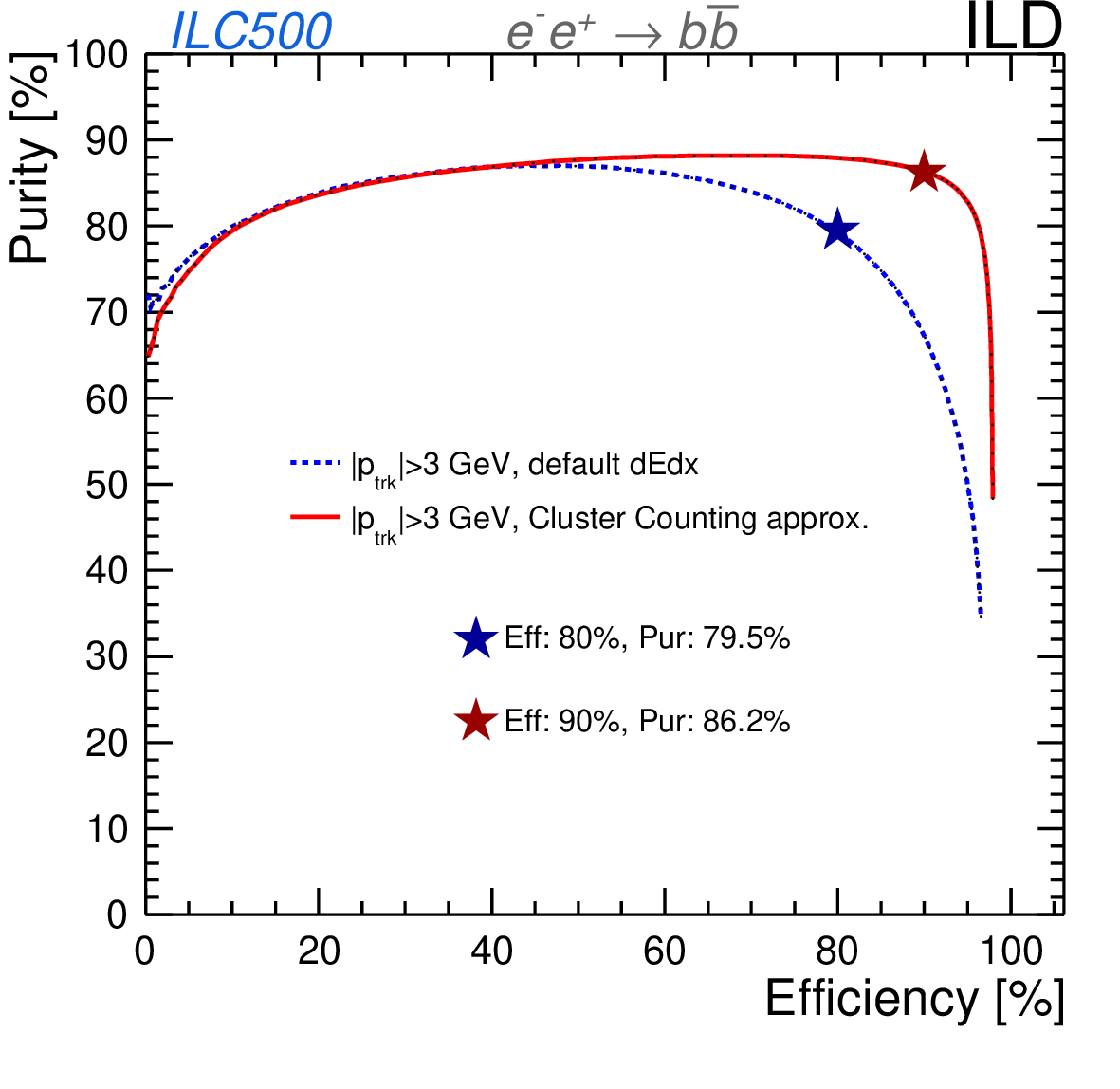} &
      \includegraphics[width=0.45\textwidth]{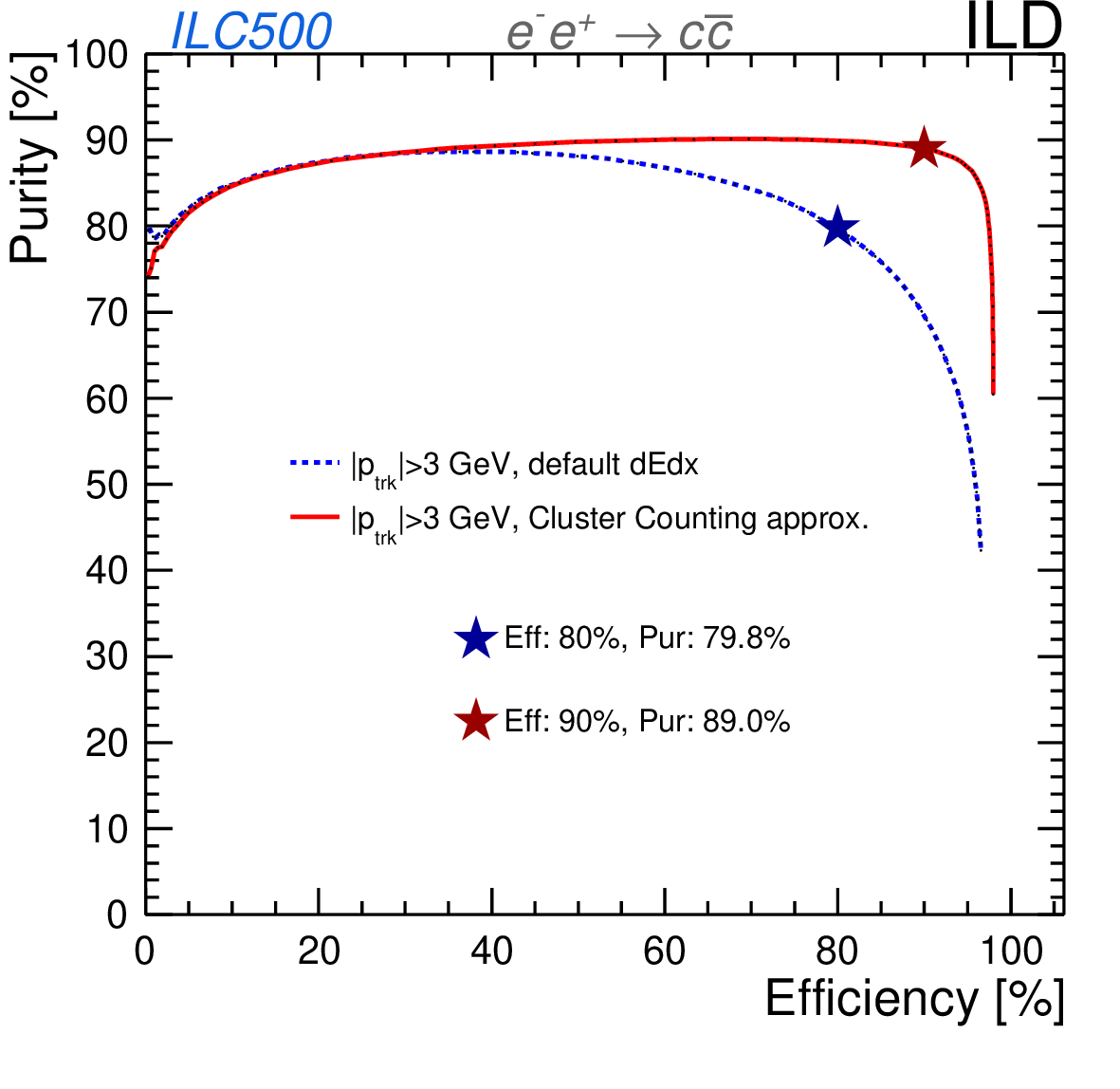} 
    \end{tabular}
\caption{Prospects on charged kaon selection (purity vs efficiency) comparing the use of \dNdx instead of \dEdx. On top row: effects in \bbbar (left) and \ccbar (right) events, both at 250 GeV. The second row shows the result for 500 GeV.  \label{fig:kaonID}}
\end{center}
\end{figure}

Once the expected improvement in resolution for the \dNdx case is accounted for, the full analysis is repeated. Given that now the \textit{likeness} distributions offer a better separation between types of hadrons (see Fig.\ref{fig:dEdxdist_cquark}), the criteria to accept a track as kaon candidate requires an absolute value of the \textit{kaonness} smaller than 1.0 instead of 1.5. Besides the separation criteria, the optimisation process is the same, including the PSO for each possible scenario. In Fig.\ref{fig:flavourtagging}, the effects on flavour tagging when adding the three new variables from kaon, protons and pions are shown for the case of \dEdx,\dNdx and ideal PID with 100\% efficiency for both 250 and 500 GeV. In Fig.\ref{fig:kaonID}, the impact of moving from \dEdx to \dNdx, envisioning a pixel PID, is shown for 250 and 500 GeV.

\section{Experimental prospects for \Afb  and \Rq at ILC250 and ILC500}
\label{sec:results}

\begin{figure}[!ht]
\begin{center}
    \begin{tabular}{cc}
      \includegraphics[width=0.45\textwidth]{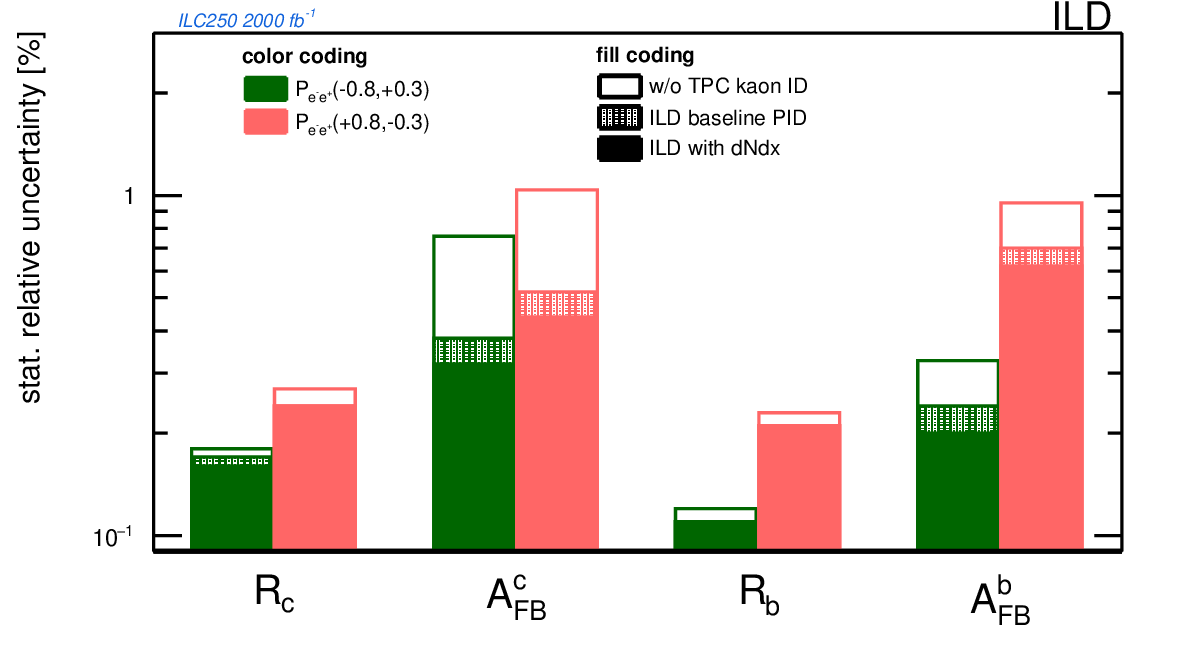} &
      \includegraphics[width=0.45\textwidth]{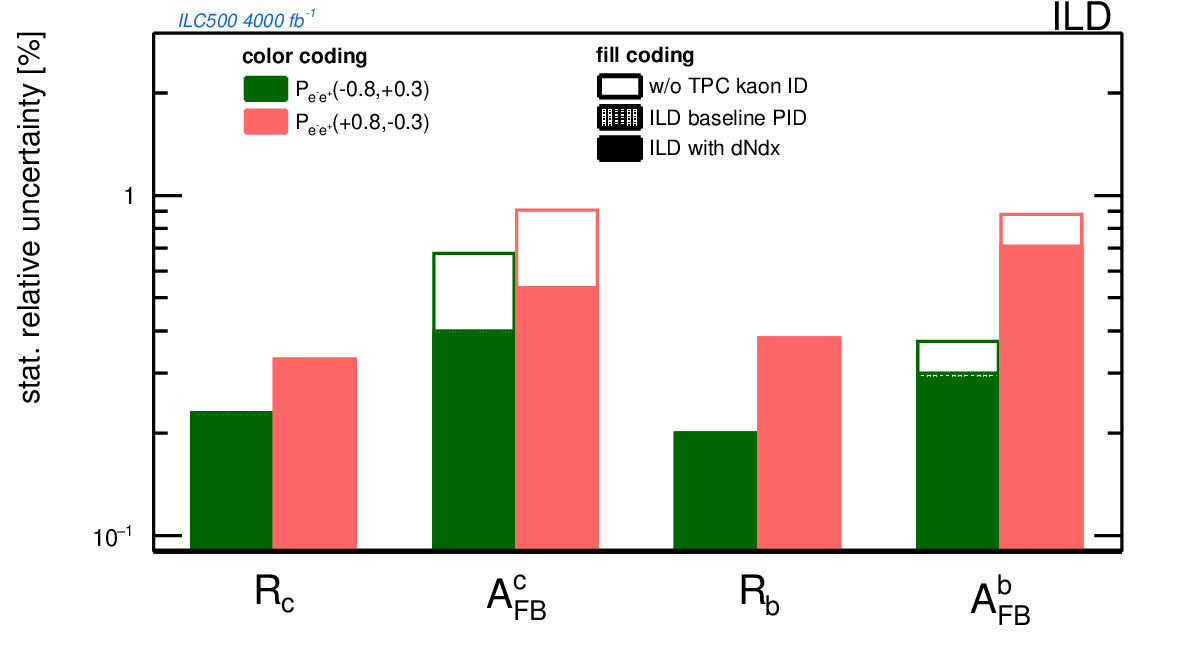} \\
      \includegraphics[width=0.45\textwidth]{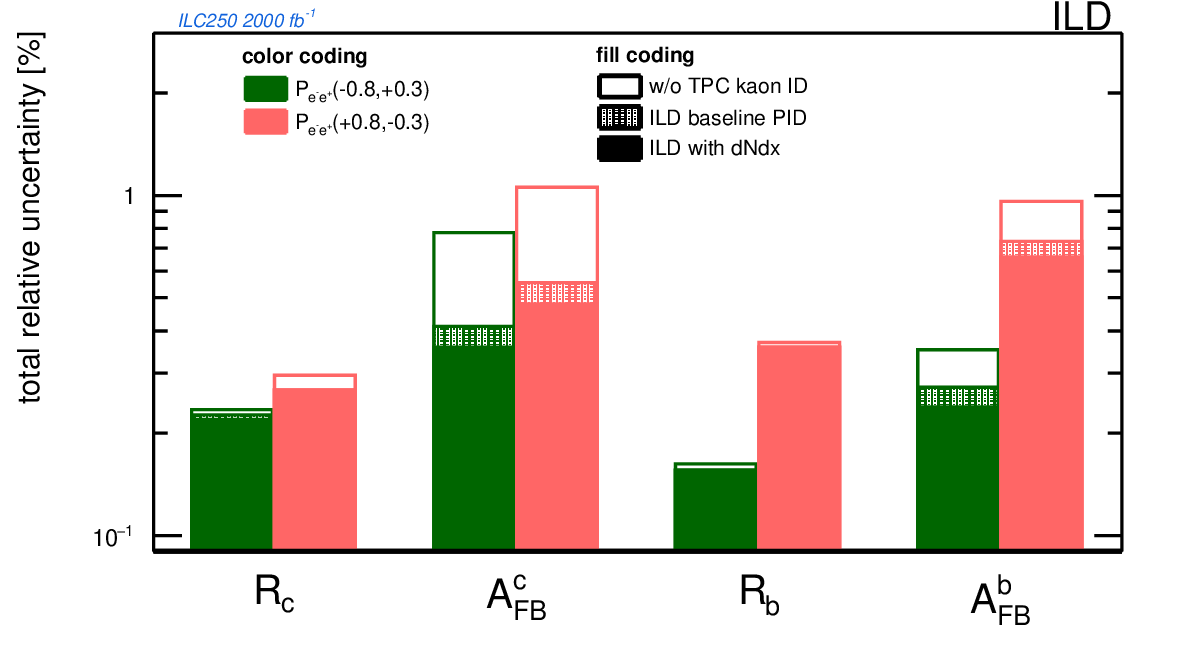} &
      \includegraphics[width=0.45\textwidth]{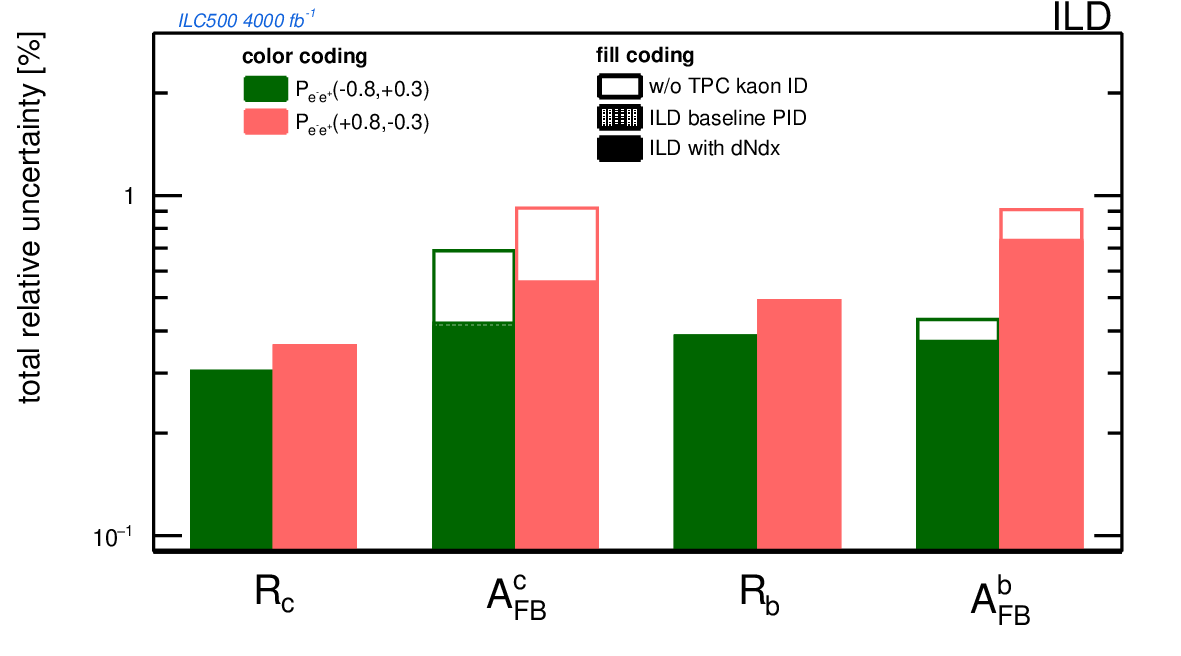} 
    \end{tabular}
\caption{Prospects on \AFB and \Rq experimental measurements at ILC250 (left column) and ILC500 (right column). In the first row, the estimations assume only statistical uncertainties. In the second row, the systematic uncertainties are also included. \label{fig:results}}
\end{center}
\end{figure}

Three different scenarios have been studied: reconstruction without TPC kaon ID, a reconstruction using TPC Kaon ID (via \dEdx) for charge measurement as well as adding \dEdx in the flavour tagging and a reconstruction using TPC Kaon ID (via \dNdx) for charge measurement as well as adding \dNdx in the flavour tagging, being the later an estimation as described in section \ref{dNdxsection}. These three scenarios are covered for both 250 and 500 GeV, for the cases of $P_{\mathrm{e^{-}e^{+}}}=(-0.8,+0.3)$ and $P_{\mathrm{e^{-}e^{+}}}=(+0.8,-0.3)$. A comparison for each case using only statistical uncertainties has also been plotted. The results are summarised in Fig. \ref{fig:results}.

\section{Summary and prospects}
\label{sec:conclusions}
In this document, an update of the previous work at ILC250 \cite{Irles:2023ojs} has been done, extending it to ILC500. 
Furthermore, a detailed study of the impact of the TPC PID capabilities at ILD in both scenarios has been covered, including prospects of using cluster counting (\dNdx) with a pixel TPC.
Adding PID from TPC measurements into the flavour tagging software for $b$ and \cquark only provides a residual improvement since the main flavour-tagging variables are related to the vertex reconstruction. However, the TPC PID plays a key role in the reconstruction of \AFB. Prospects of a pixel TPC using \dNdx for PID envision a noticeable impact in the charge reconstruction of the \qqbar systems, especially for the \cquark case, where a factor $\sim2$ improvement in the experimental precision is obtained.

Ongoing work by the authors of this note is the evaluation of experimental sensitivity of BSM models predicting new heavy resonances $Z^{\prime}$\cite{Yoon:2018xud,Funatsu:2017nfm,Funatsu:2020haj}, which could affect the predictions of \Rq and \AFB at ILC250/500. 

The ILC program also envisions a run at the Z-pole, the GigaZ mode\cite{irles2019complementarity}. 
With that run. one would expect at least one order of magnitude of improvements for the Z-couplings measurements of \bquark and \cquark compared to those of SLC and LEP and give a complete picture of the quark EW couplings when combining measurements at Giga-Z, ILC250 and ILC500. However, since a study at the Giga-Z would involve the development of new simulations and methodologies, it is considered out of this work's scope. 


\section*{Acknowledgements}
We would like to thank the LCC generator working group and the ILD software working group for providing the simulation and reconstruction tools and producing the Monte Carlo samples used in this study.
This work has benefited from computing services provided by the ILC Virtual Organization, supported by the national resource providers of the EGI Federation and the Open Science GRID.

The authors are funded by the Generalitat Valenciana (Spain) under the grant number CIDEGENT/2020/21. AI also acknowledges the financial support from the MCIN with funding from the European Union NextGenerationEU and Generalitat Valenciana in the call Programa de Planes Complementarios de I+D+i (PRTR 2022) Project \textit{Si4HiggsFactories}, reference ASFAE$/2022/015$.

\section{Appendix}

\subsection{Flavour tagging variables}
\label{sec:annex}
\begin{table}[ht]
\centering
\begin{tabularx}{\linewidth}{|c|L|}
\hline
Cat. & Variables                                                                                  \\ \hline
A    & trk1d0sig trk2d0sig trk1z0sig trk2z0sig trk1pt\_jete trk2pt\_jete jprobr2 jprobr25sigma jprobz2 jprobz25sigma d0bprob2 d0cprob2 d0qprob2 z0bprob2 z0cprob2 z0qprob2 \textbf{dEdxNKaonSec} \textbf{dEdxNPionSec} \textbf{dEdxNProtonSec} \\ \hline
B    & trk1d0sig trk2d0sig trk1z0sig trk2z0sig trk1pt\_jete trk2pt\_jete jprobr2 jprobz2
      vtxlen1\_jete vtxsig1\_jete vtxdirang1\_jete vtxmom1\_jete vtxmass1 vtxmult1 vtxmasspc vtxprob
      d0bprob2 d0cprob2 d0qprob2 z0bprob2 z0cprob2 z0qprob2
      trkmass nelectron nmuon
      \textbf{dEdxNKaonSec} \textbf{dEdxNPionSec} \textbf{dEdxNProtonSec}  \\ \hline
C    &   trk1d0sig trk2d0sig trk1z0sig trk2z0sig trk1pt\_jete trk2pt\_jete jprobr2 jprobz2
      vtxlen1\_jete vtxsig1\_jete vtxdirang1\_jete vtxmom1\_jete vtxmass1 vtxmult1 vtxmasspc vtxprob
      1vtxprob vtxlen12all\_jete vtxmassall
      \textbf{dEdxNKaonSec} \textbf{dEdxNPionSec} \textbf{dEdxNProtonSec} \\ \hline
D    & trk1d0sig trk2d0sig trk1z0sig trk2z0sig trk1pt\_jete trk2pt\_jete jprobr2 jprobz2
      vtxlen1\_jete vtxsig1\_jete vtxdirang1\_jete vtxmom1\_jete vtxmass1 vtxmult1 vtxmasspc vtxprob
      vtxlen2\_jete vtxsig2\_jete vtxdirang2\_jete vtxmom2\_jete vtxmass2 vtxmult2
      vtxlen12\_jete vtxsig12\_jete vtxdirang12\_jete vtxmom\_jete vtxmass vtxmult
      1vtxprob 
      \textbf{dEdxNKaonSec} \textbf{dEdxNPionSec} \textbf{dEdxNProtonSec} \\ \hline
\end{tabularx}
\caption{Variables used in the TMVA's BDT in each of the categories in \texttt{LCFI+}. This selection of variables is the same for the case with \dEdx, \dNdx or ideal PID scenarios. The bold text represent the new variables while the rest are original \texttt{LCFI+} variables introduced in \cite{Suehara:2015ura}.
\label{tab:bdtvariables}}
\end{table}

\newpage

\subsection{Reconstruction performance plots for the  ILC500 case}
\label{sec:appendixPlots}

\begin{figure}[!ht]
\begin{center}
    \begin{tabular}{cc}
      \includegraphics[width=0.45\textwidth]{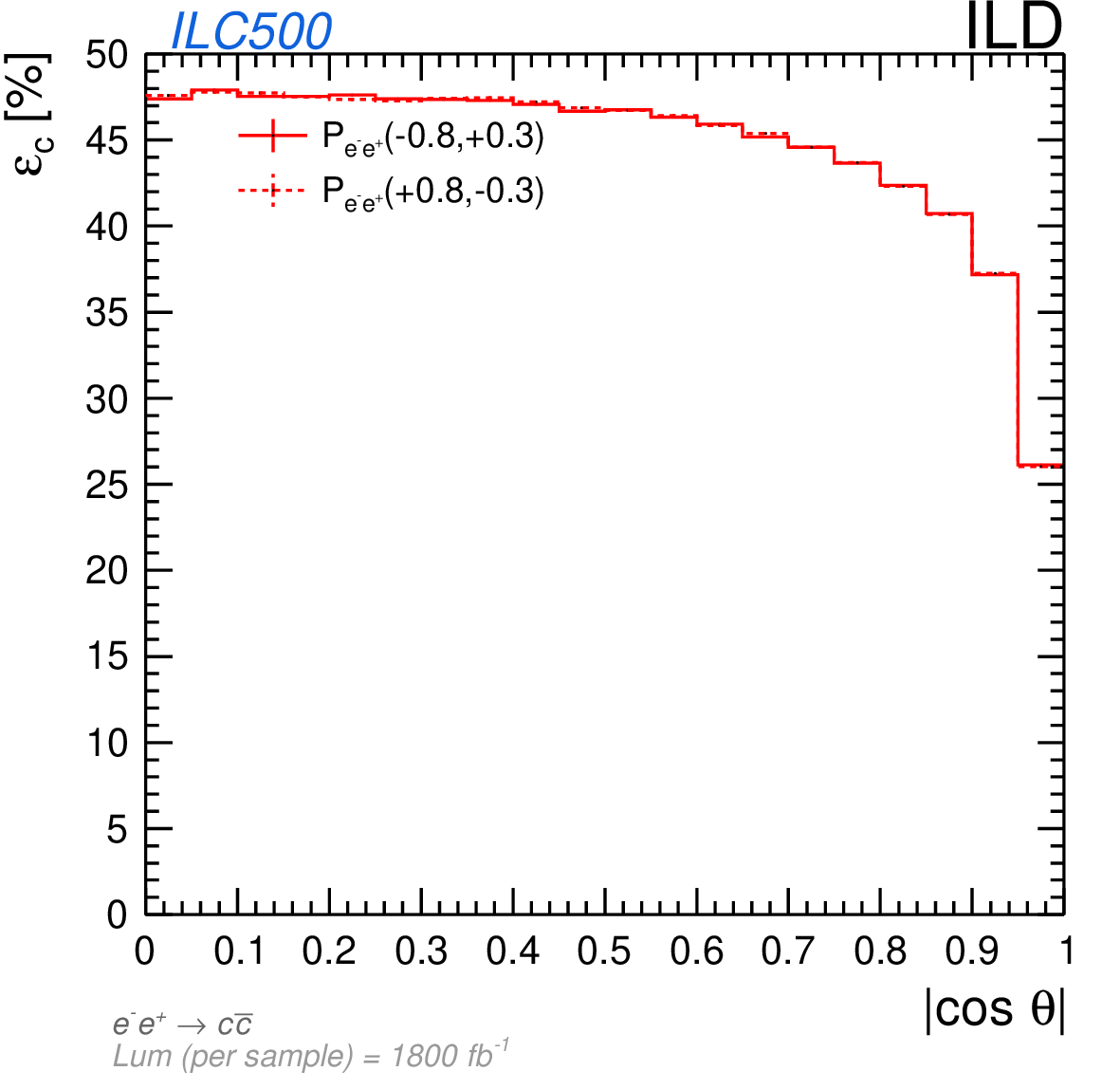} & \includegraphics[width=0.45\textwidth]{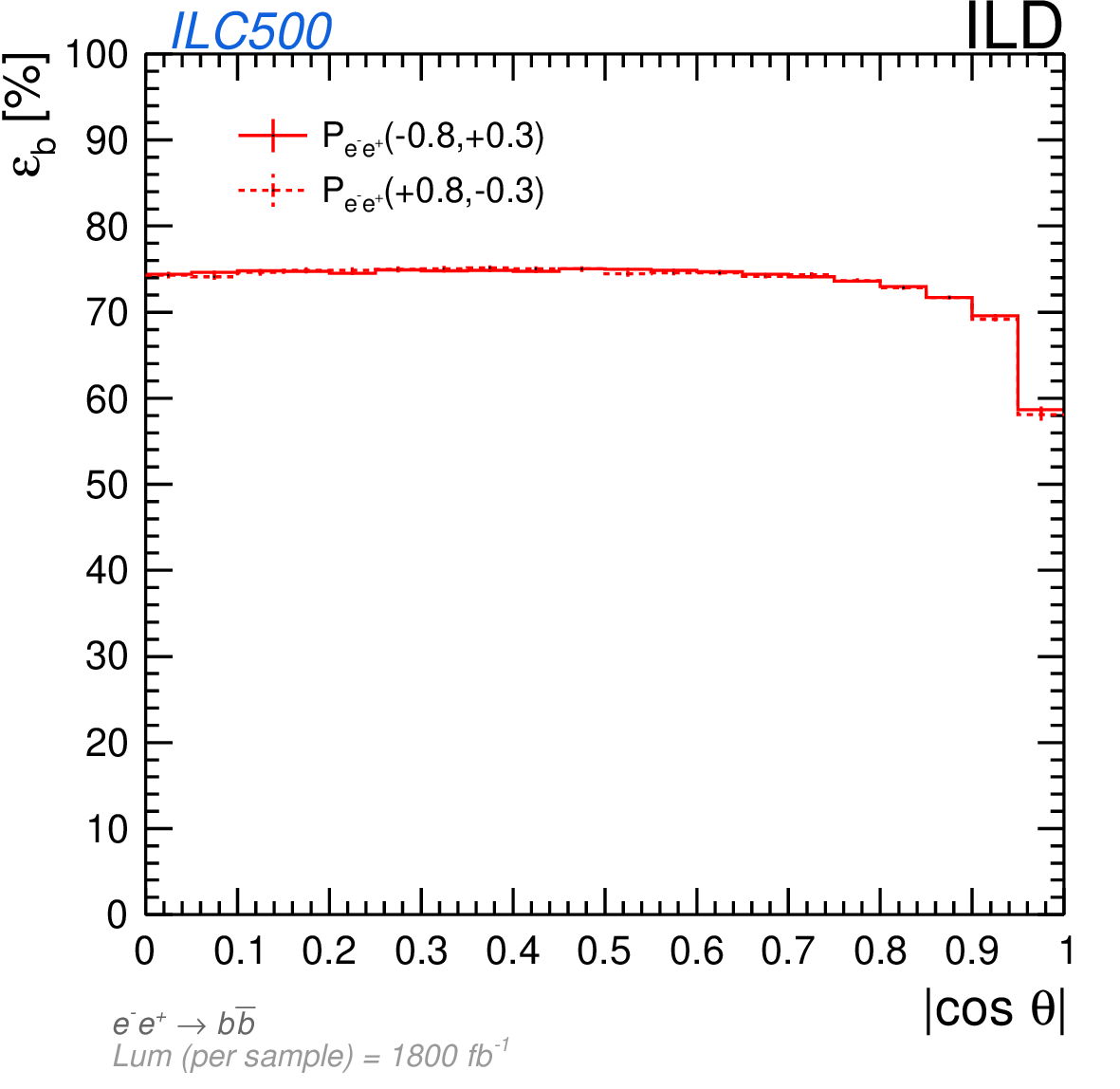} \\
      \includegraphics[width=0.45\textwidth]{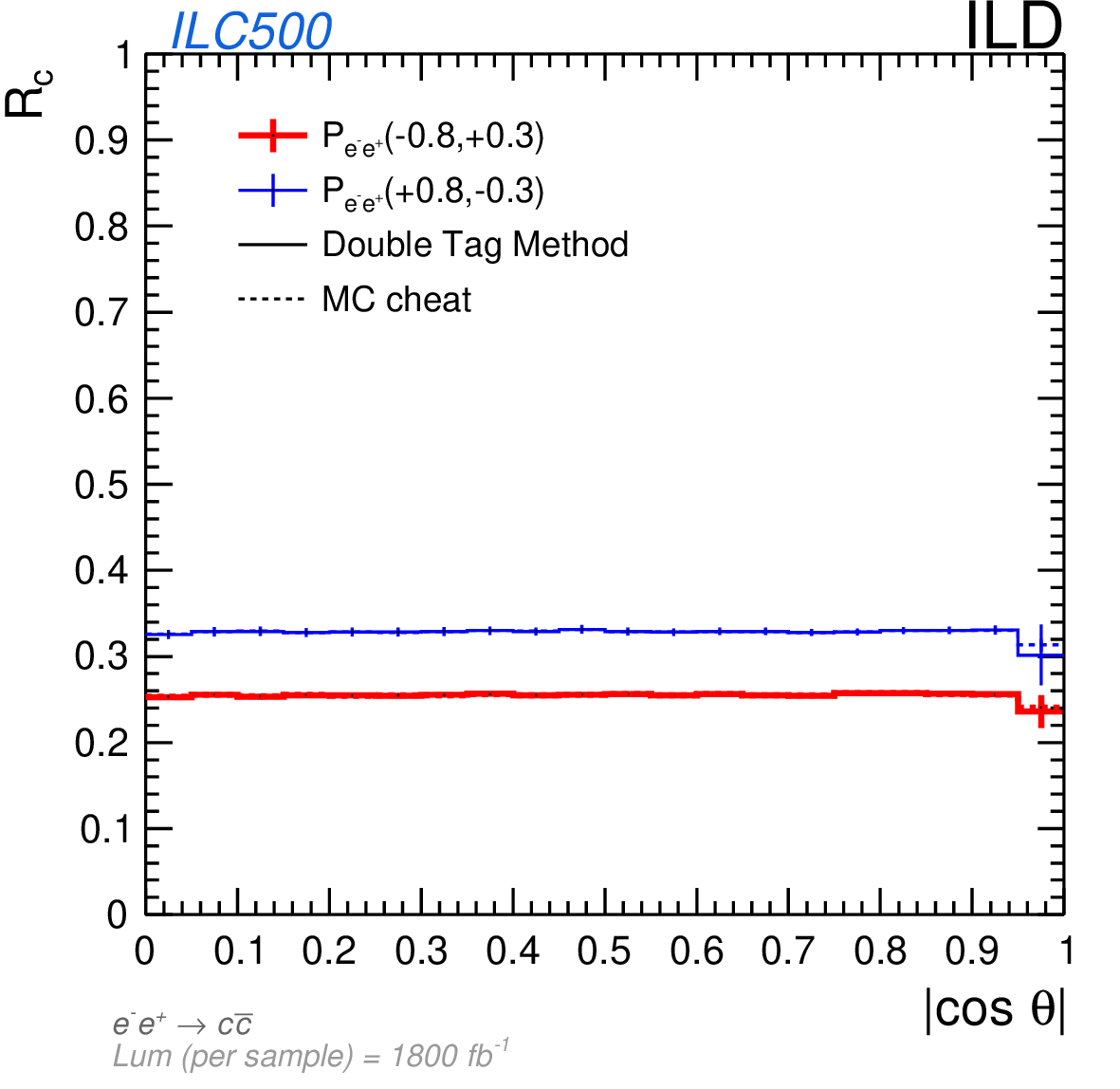} & \includegraphics[width=0.45\textwidth]{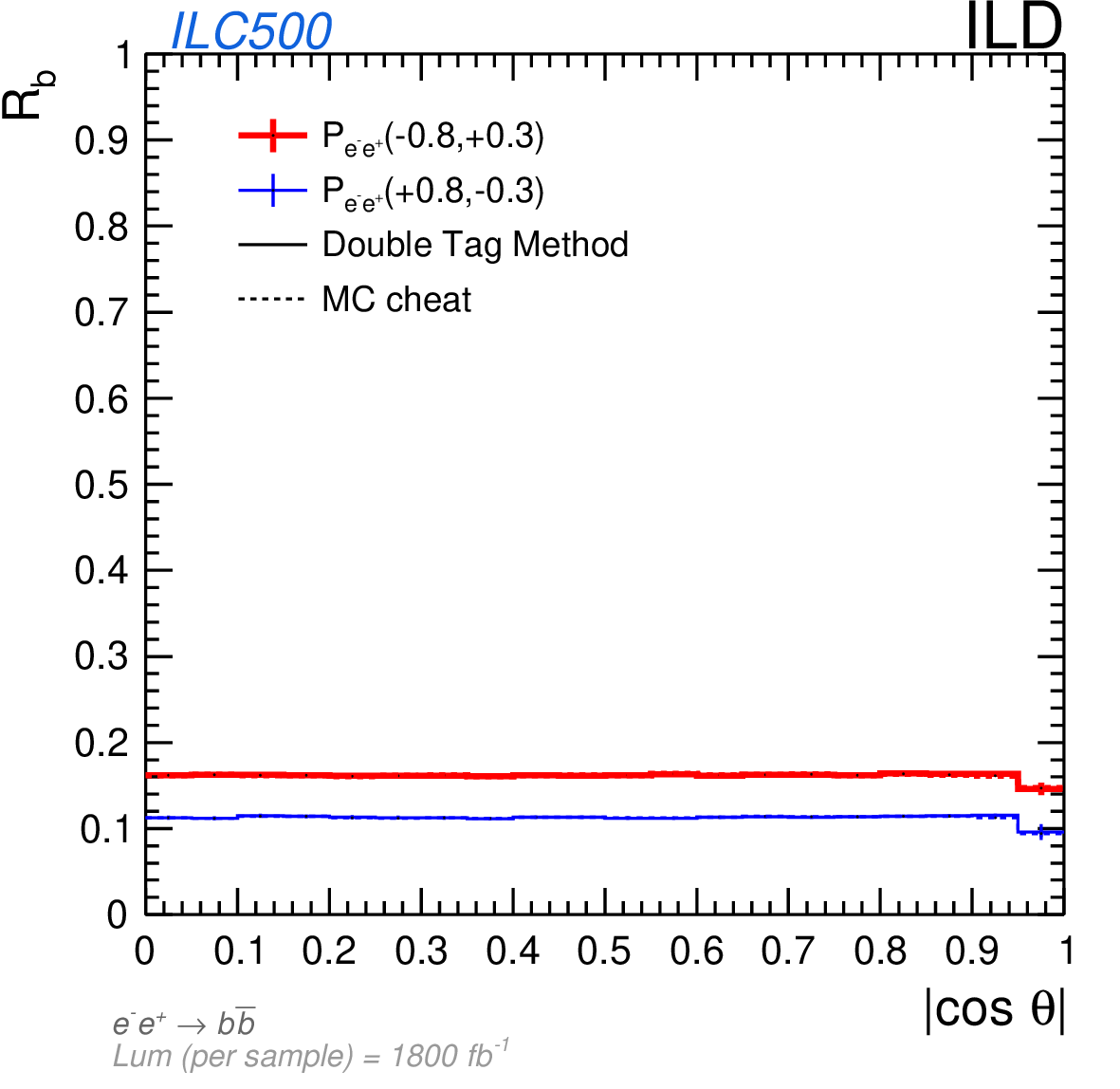}
    \end{tabular}
\caption{Extracted $\epsilon_{q}$ and \Rq using the DT method for \eecc (left) and \eebb (right). For the estimation of \Rq the comparison with the case in which the Monte Carlo is used to estimate the $\epsilon_{q}$ is included, showing no difference with the DT expectations.
\label{fig:RbRc}}
\end{center}
\end{figure}

\begin{figure}[!ht]
\begin{center}
    \begin{tabular}{cc}
      \includegraphics[width=0.45\textwidth]{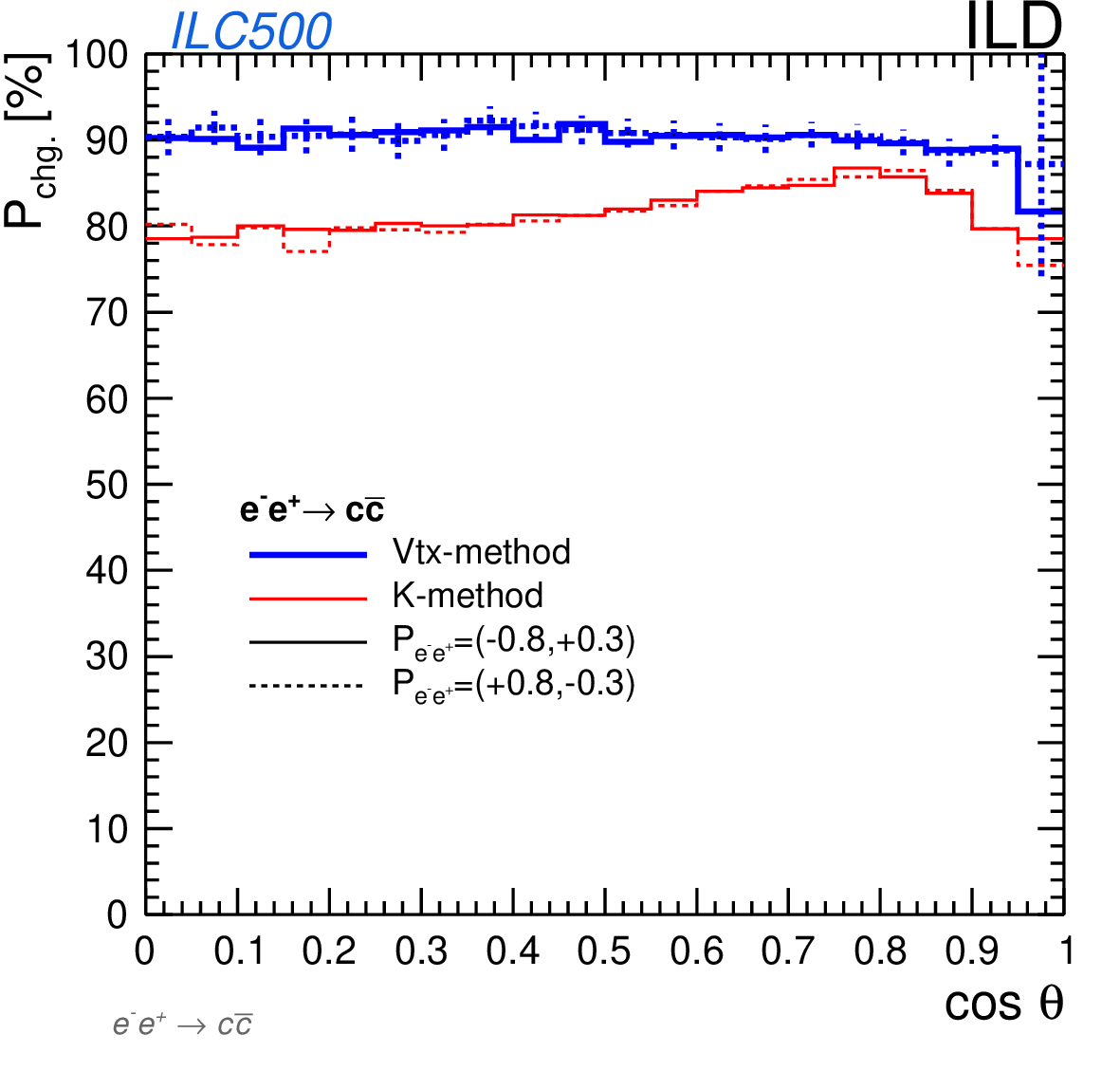} &
      \includegraphics[width=0.45\textwidth]{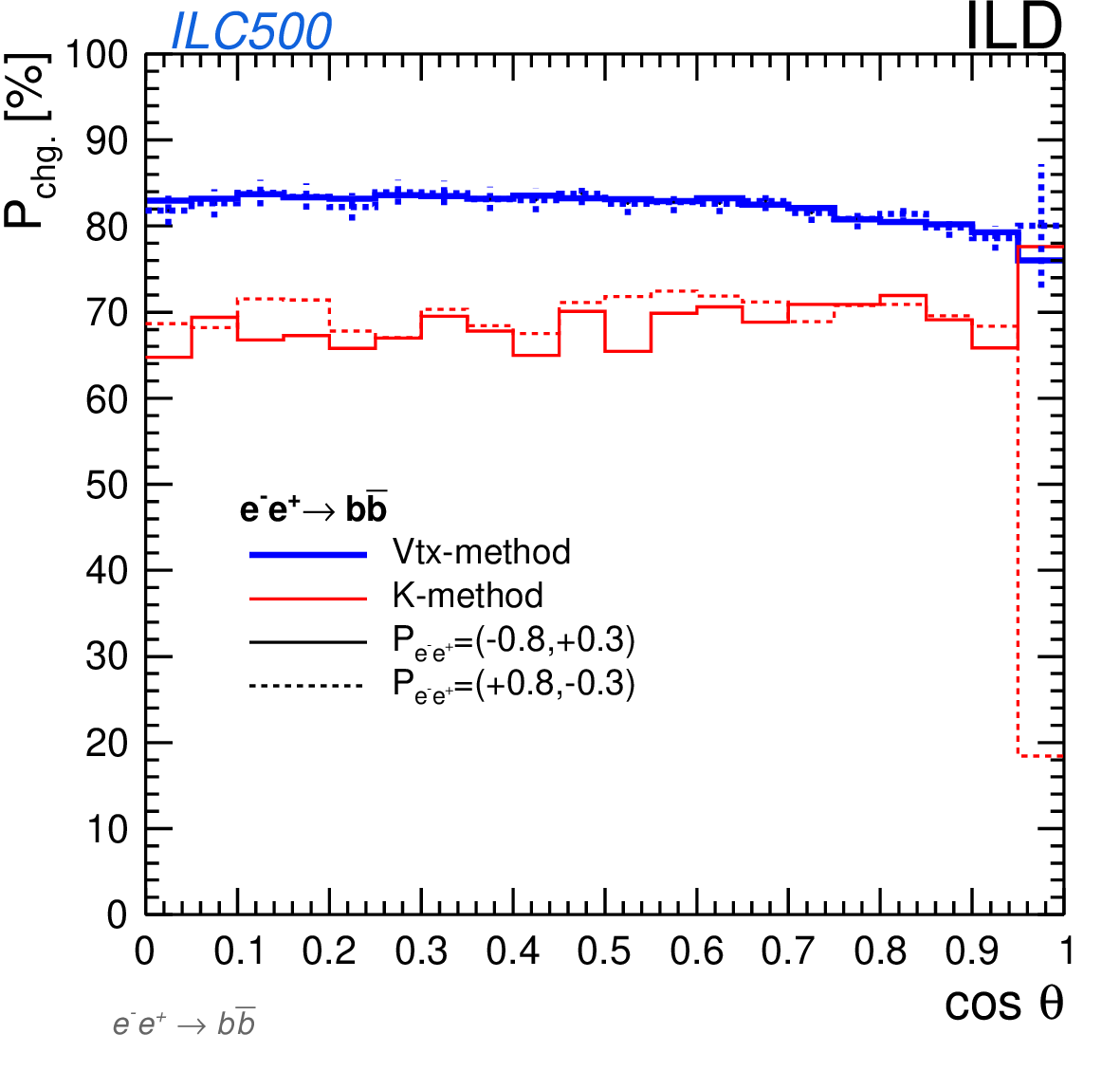} 
    \end{tabular}
\caption{Distributions for \Pb for the \Bc (blue) and the \Kc (red) for \ccbar (left) and \bbbar (right). The results are shown for different beam polarisation scenarios using different types of lines.  This figure is the correspondence to Figure 15 from \cite{Irles:2023ojs} but for ILC500 and using the \dEdx information also for the flavour tagging.
\label{fig:purity}}
\end{center}
\end{figure}

\begin{figure}[!ht]
\begin{center}
    \begin{tabular}{cc}
      \includegraphics[width=0.45\textwidth]{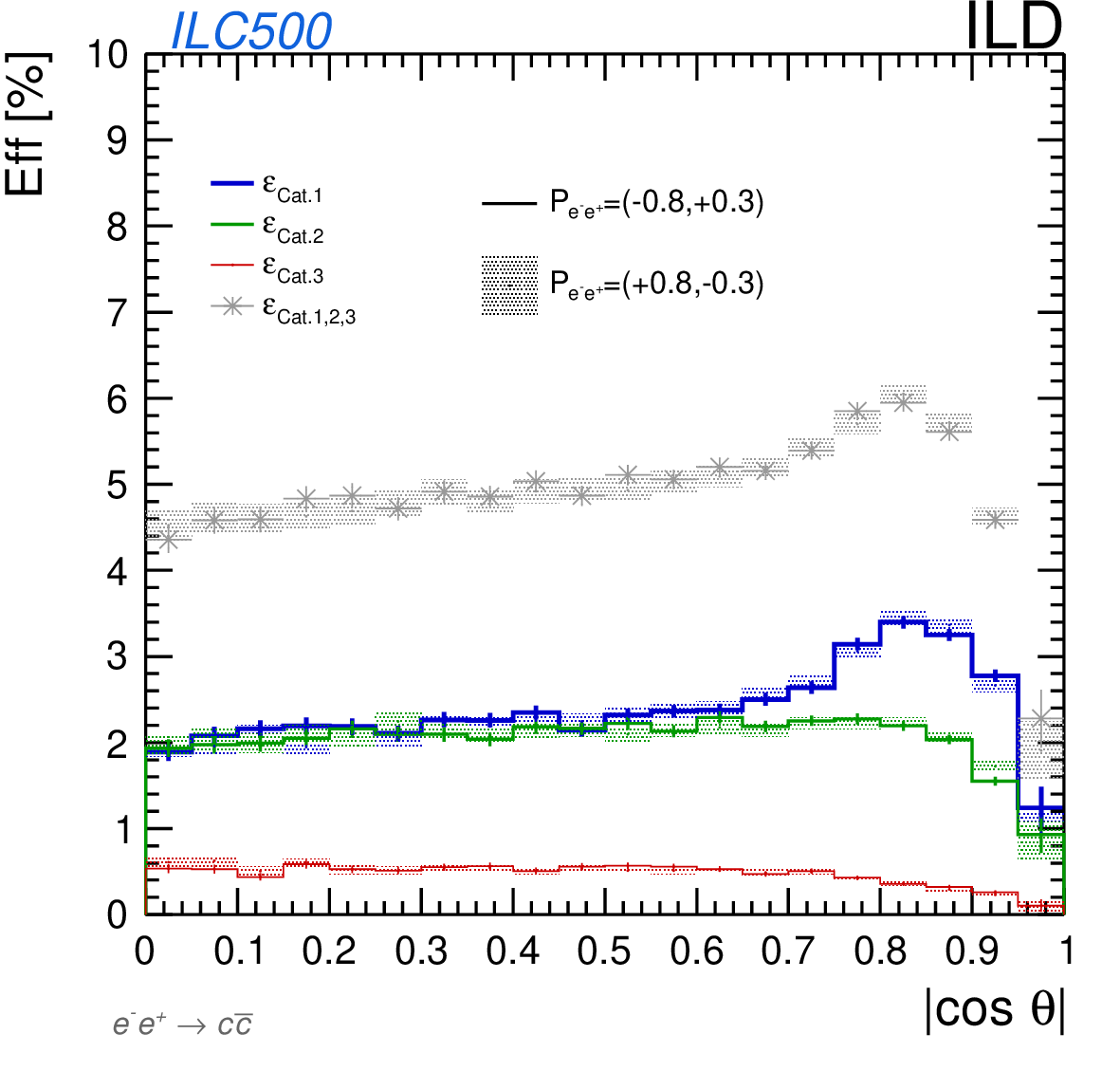} &
      \includegraphics[width=0.45\textwidth]{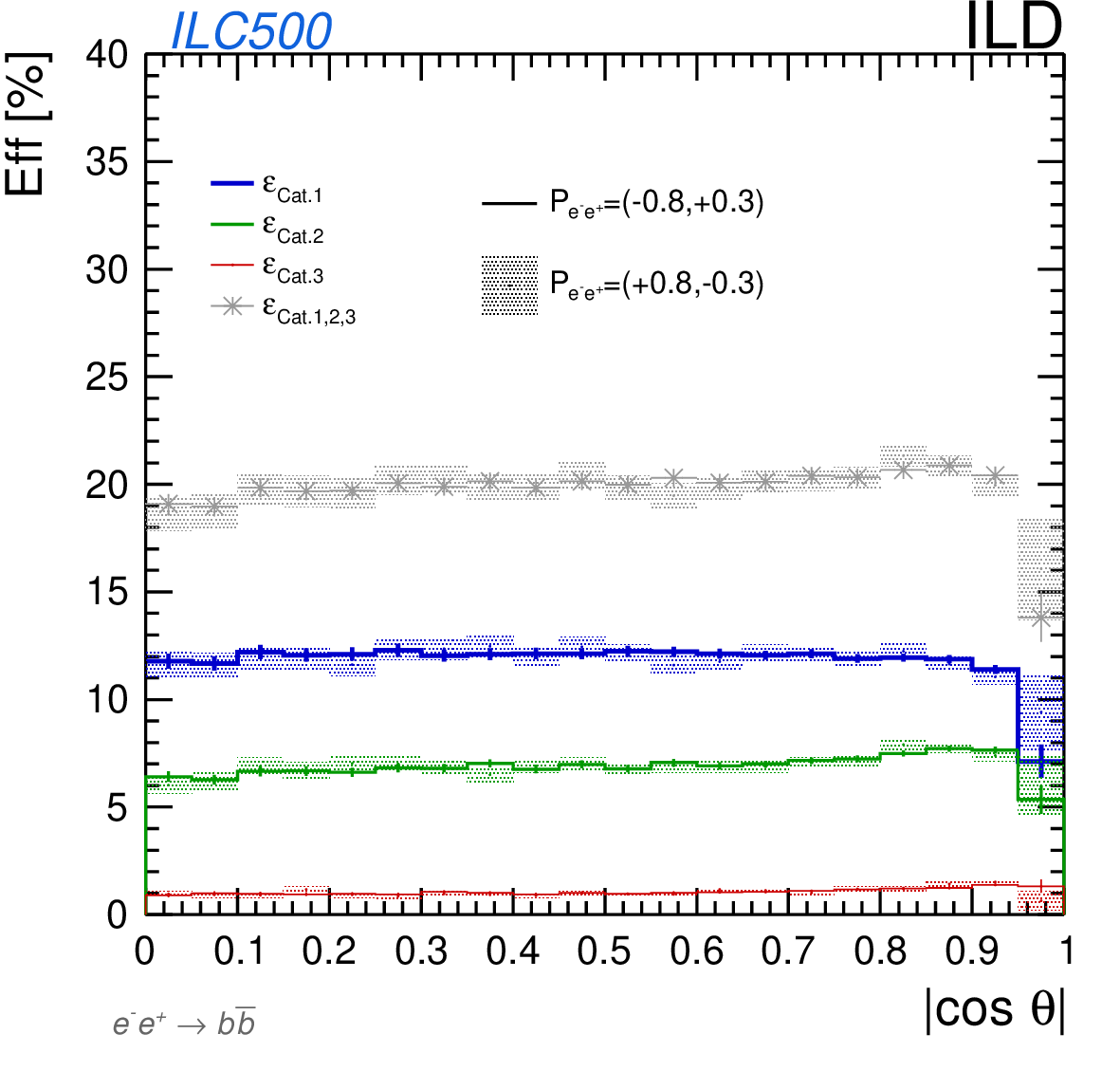} 
    \end{tabular}
      \caption{Distribution of the different selection efficiencies for \ccbar (left) and \bbbar (right) for the \AFB measurement, as described in Eq. 20-22 \cite{Irles:2023ojs} For the \ccbar (\bbbar) case, the $Cat.1$ corresponds to only the \Kc (\Bc) applied and $Cat.2$ to only the \Bc (\Kc) applied. This figure is the correspondence to Figure 18 from \cite{Irles:2023ojs} but for ILC500 and using the \dEdx information also for the flavour tagging. \label{fig:eff_cat}}
\end{center}
\end{figure}

\begin{figure}[!ht]
\begin{center}
    \begin{tabular}{cc}
      \includegraphics[width=0.45\textwidth]{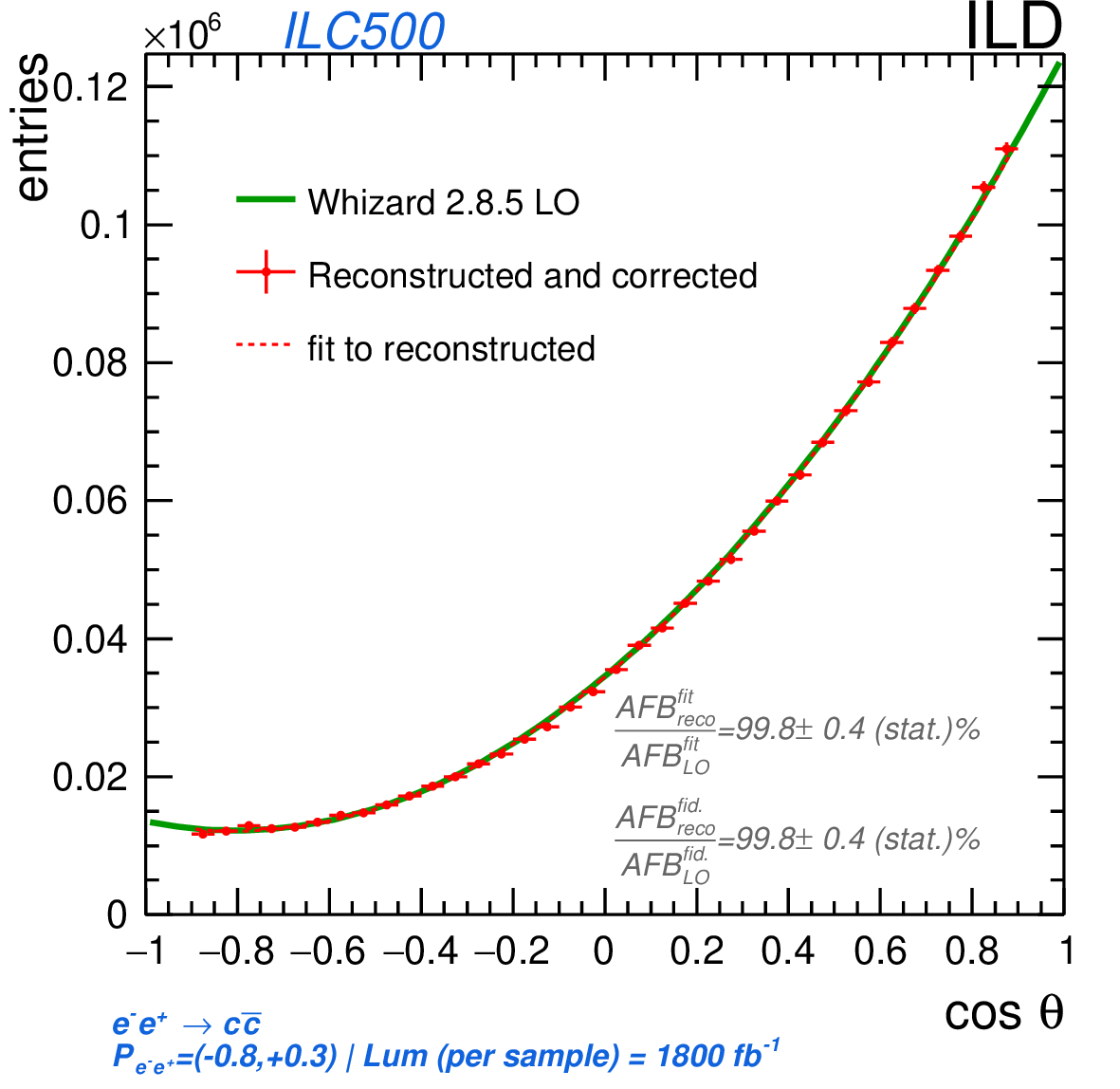} &
      \includegraphics[width=0.45\textwidth]{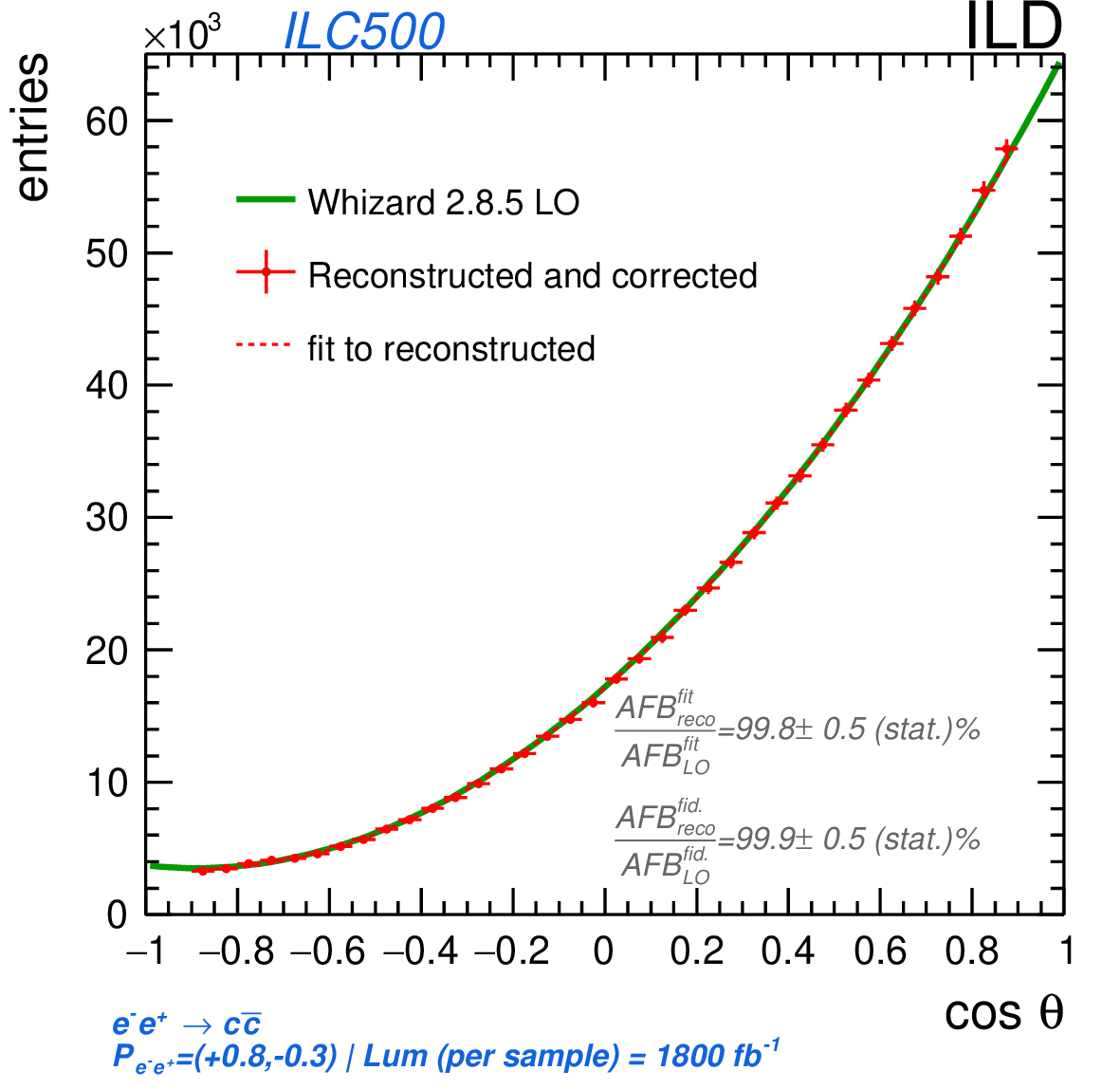} \\
      \includegraphics[width=0.45\textwidth]{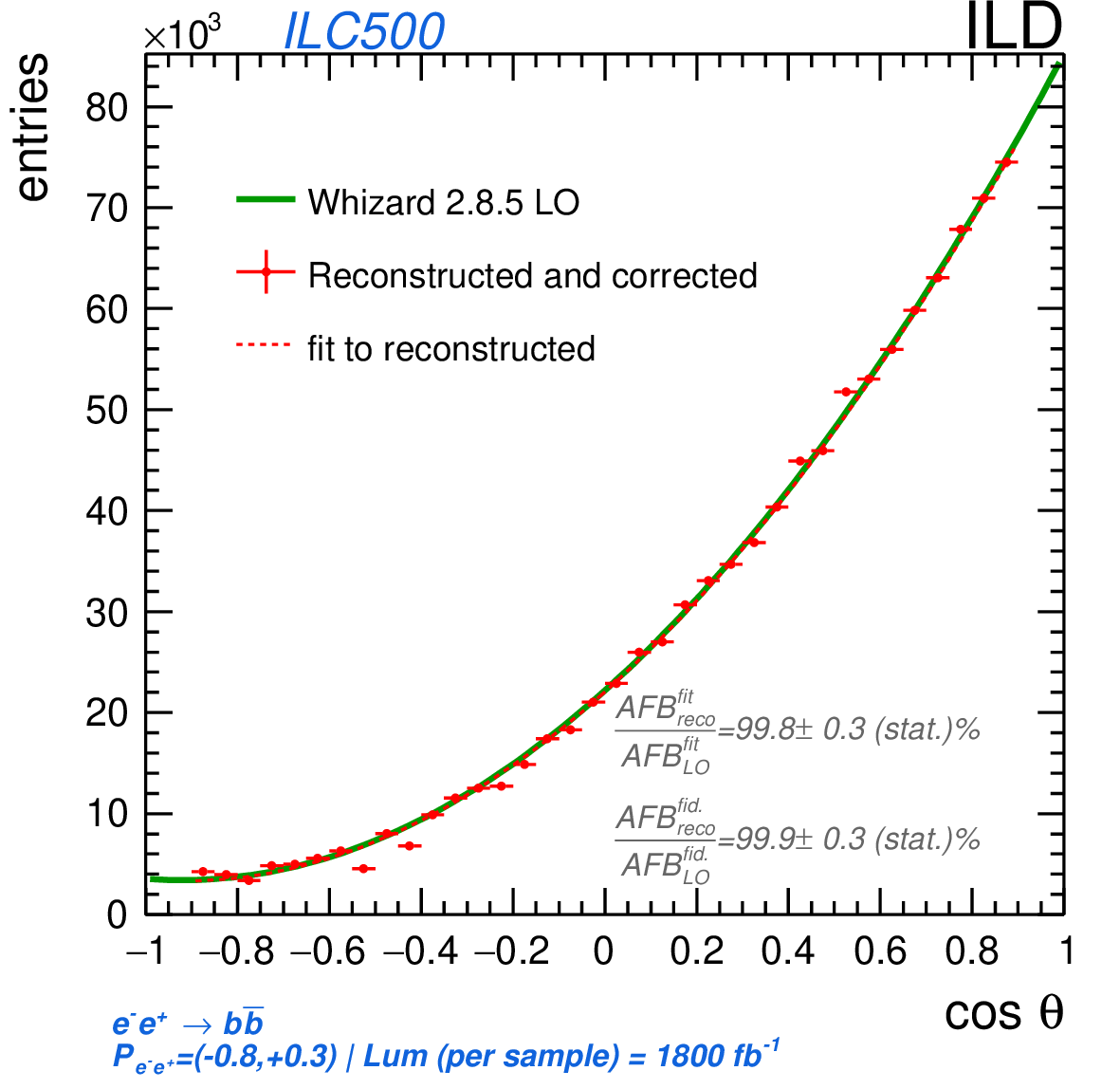} &
      \includegraphics[width=0.45\textwidth]{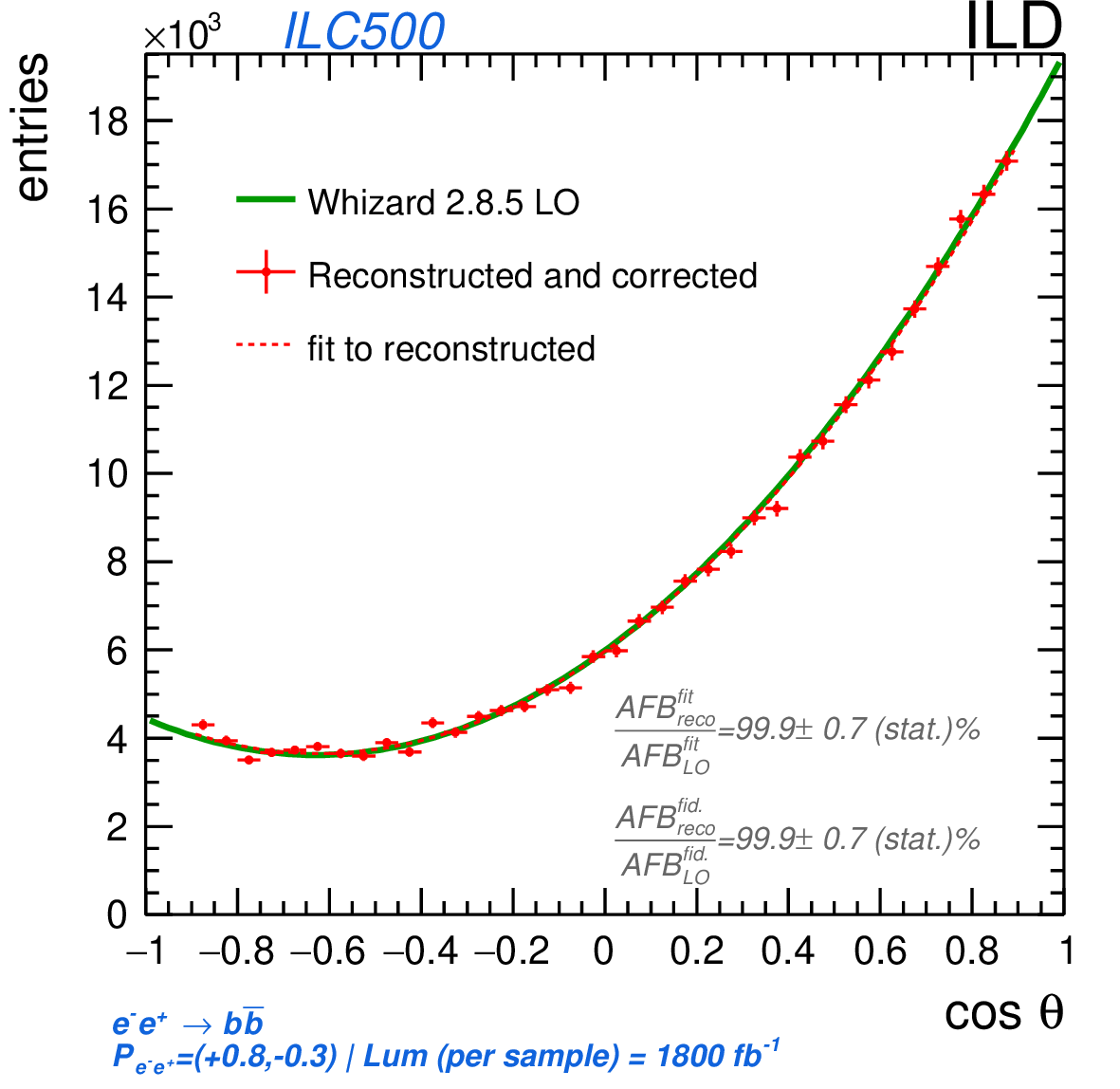} 
    \end{tabular}
      \caption{Fit (red graph) of the final distributions (red points) to the function described in Equation 23 from \cite{Irles:2023ojs}
      The fit is performed between $-0.9<\costheta<0.9$ to avoid the regions with a large correction due to efficiency and acceptance losses.
      The result of the fit is extended to $[-1,1]$ for the \AFB estimation.
      The LO calculations are represented in the green graphs. The forward-backward asymmetry after the full reconstruction
      and the LO prediction are compared and are well in agreement with statistical uncertainties. This figure is the correspondence to Figure 19 from \cite{Irles:2023ojs} but for ILC500 and using the \dEdx information also for the flavour tagging. \label{fig:AFB_fit}}
\end{center}
\end{figure}

\newpage%
\clearpage

\printbibliography[title=References]
\end{document}